\documentclass[fleqn,usenatbib]{mnras}

\usepackage[T1]{fontenc}

\DeclareRobustCommand{\VAN}[3]{#2}
\let\VANthebibliography\thebibliography
\def\thebibliography{\DeclareRobustCommand{\VAN}[3]{##3}\VANthebibliography}

\usepackage{graphicx}	
\usepackage{amsmath}	

\usepackage{newtxtext,newtxmath}

\newcommand\Msun{\,\rmn{M}_{\sun}}

\newcommand\Gyr{\,\rmn{Gyr}}

\newcommand\kpc{\,\rmn{kpc}}
\newcommand\Mpc{\,\rmn{Mpc}}
\newcommand\MLsun{\,(\rmn{M/L})_{\sun}}

\newcommand{\kms}{\,\rmn{km}\,\rmn{s}^{-1}}
\newcommand{\ts}{\textsuperscript}

\newcommand{\gadget}   {\textsc{gadget3}}
\newcommand{\subfind}  {\textsc{subfind}}

\newcommand{\K}        {\,{\rm K}}
\newcommand{\cmcubed}  {\,{\rm cm}^{-3}}

\newcommand{\Mcstar}   {M_\rmn{c,\ast}}

\title[UDG kinematics and GC system richness]{Origin of the correlation between stellar kinematics and globular cluster system richness in ultra-diffuse galaxies}

\author[J. Pfeffer et al.]{Joel Pfeffer,$^{1,2}$\thanks{E-mail: jpfeffer@swin.edu.au (JP)}
Steven R.~Janssens,$^{1,2}$
Maria Luisa Buzzo,$^{1,2}$
Jonah S.~Gannon,$^{1,2}$
Nate Bastian,$^{3,4}$\newauthor
Kenji Bekki,$^{5}$
Jean P.~Brodie,$^{1,2,6}$
Warrick J.~Couch,$^{1}$
Robert A.~Crain,$^{7}$
Duncan A.~Forbes,$^{1,2}$\newauthor
J.~M.~Diederik Kruijssen,$^{8,9}$
Aaron J.~Romanowsky$^{10,11}$
\\
$^{1}$Centre for Astrophysics \& Supercomputing, Swinburne University, Hawthorn, VIC 3122, Australia\\
$^{2}$ARC Centre of Excellence for All Sky Astrophysics in 3 Dimensions (ASTRO 3D), Australia\\
$^{3}$Donostia International Physics Center (DIPC), Paseo Manuel de Lardizabal, 4, E-20018 Donostia-San Sebasti\'{a}n, Guipuzkoa, Spain\\
$^{4}$IKERBASQUE, Basque Foundation for Science, E-48013 Bilbao, Spain\\
$^{5}$International Centre for Radio Astronomy Research, University of Western Australia, 35 Stirling Highway, Crawley, WA 6009, Australia\\
$^{6}$University of California Observatories, 1156 High Street, Santa Cruz, CA 95064, USA\\
$^{7}$Astrophysics Research Institute, Liverpool John Moores University, 146 Brownlow Hill, Liverpool L3 5RF, UK\\
$^{8}$Technical University of Munich, School of Engineering and Design, Department of Aerospace and Geodesy, Chair of Remote Sensing Technology, \\\hspace{2.2mm}Arcisstr. 21, 80333 Munich, Germany\\
$^{9}$Cosmic Origins Of Life (COOL) Research DAO, coolresearch.io\\
$^{10}$Department of Physics and Astronomy, San Jos\'{e} State University, One Washington Square, San Jose, CA 95192, USA\\
$^{11}$Department of Astronomy \& Astrophysics, University of California Santa Cruz, 1156 High Street, Santa Cruz, CA 95064, USA
}

\date{Accepted 2024 March 19. Received 2024 March 01; in original form 2023 December 13}

\pubyear{2023}

\begin{document}
\label{firstpage}
\pagerange{\pageref{firstpage}--\pageref{lastpage}}
\maketitle

\begin{abstract}
Observational surveys have found that the dynamical masses of ultra-diffuse galaxies (UDGs) correlate with the richness of their globular cluster (GC) system.
This could be explained if GC-rich galaxies formed in more massive dark matter haloes.
We use simulations of galaxies and their GC systems from the E-MOSAICS project to test whether the simulations reproduce such a trend.
We find that GC-rich simulated galaxies in galaxy groups have enclosed masses that are consistent with the dynamical masses of observed GC-rich UDGs.
However, simulated GC-poor galaxies in galaxy groups have higher enclosed masses than those observed.
We argue that GC-poor UDGs with low stellar velocity dispersions are discs observed nearly face on, such that their true mass is underestimated by observations.
Using the simulations, we show that galactic star-formation conditions resulting in dispersion-supported stellar systems also leads to efficient GC formation.
Conversely, conditions leading to rotationally-supported discs leads to inefficient GC formation.
This result may explain why early-type galaxies typically have richer GC systems than late-type galaxies.
This is also supported by comparisons of stellar axis ratios and GC specific frequencies in observed dwarf galaxy samples, which show GC-rich systems are consistent with being spheroidal, while GC-poor systems are consistent with being discs.
Therefore, particularly for GC-poor galaxies, rotation should be included in dynamical mass measurements from stellar dynamics.
\end{abstract}




\section{Introduction}

Although dwarf galaxies (stellar masses $M_\ast \lesssim 10^9 \Msun$) only make up a small fraction of the present-day stellar mass density, they are the most numerous type of galactic system in the Universe \citep{Li_and_White_09}.
Their properties (e.g.\ being dark-matter dominated systems) make them strong probes of both galaxy formation and cosmological models \citep[see][for a recent review]{Sales_Wetzel_and_Fattahi_22}.

Recent years have seen considerable interest in a population of dwarf galaxies with large sizes ($R_\mathrm{eff} \gtrsim 1.5 \kpc$) and low central surface brightness ($\mu_\mathrm{g,0} \gtrsim 24$~mag~arcsec$^{-2}$), typically termed ultra-diffuse galaxies \citep[UDGs,][]{van_Dokkum_et_al_15a}.
Since their identification in the Coma cluster \citep{van_Dokkum_et_al_15a, van_Dokkum_et_al_15b}, UDGs have also been discovered in other clusters, galaxy groups and field environments \citep[e.g.][]{Martinez-Delgado_et_al_16, van_der_Burg_et_al_16, Janssens_et_al_17, Leisman_et_al_17, Roman_and_Trujillo_17, La_Marca_et_al_22, Zaritsky_et_al_23}.
Their half-light radii are considerably larger than other early-type dwarf galaxies with similar stellar masses, which typically have sizes $R_\mathrm{eff} \lesssim 1 \kpc$ \citep[e.g.][]{Misgeld_and_Hilker_11, McConnachie_12}, though some late-type galaxies of similar mass have comparable sizes, \citep{Hunter_and_Elmegreen_06, Baldry_et_al_12}. 
Thus, the stellar dynamics of UDGs can probe dynamical masses to larger physical radii compared to `classical' early-type dwarf galaxies of similar stellar mass (approaching radii reached by HI discs), enabling stronger tests of the dark matter content of such gas-free galaxies.

Kinematic measurements, using both stellar and globular cluster system dynamics, have revealed a wide range in the dark matter mass content of UDGs, from dark matter-dominated systems \citep{Beasley_et_al_16, van_Dokkum_et_al_16, van_Dokkum_et_al_19, Toloba_et_al_18, Toloba_et_al_23, Martin-Navarro_et_al_19, Gannon_et_al_20, Gannon_et_al_21, Gannon_et_al_22, Gannon_et_al_23, Forbes_et_al_21, Toloba_et_al_23} to dark matter-deficient systems \citep[NGC1052-DF2 and NGC1052-DF4,][]{van_Dokkum_et_al_18a, van_Dokkum_et_al_19_DF4, Danieli_et_al_19, Emsellem_et_al_19, Muller_et_al_20, Shen_et_al_23}.
Similarly, for field UDGs with HI discs, measurements of their rotation curves yield estimates from them residing in normal dwarf galaxy-mass haloes \citep[halo virial masses $M_{200} \approx 10^{10.5}$-$10^{11} \Msun$,][]{Leisman_et_al_17, Shi_et_al_21} to being baryon-dominated systems \citep{Mancera_Pina_et_al_19, Mancera_Pina_et_al_22, Kong_et_al_22}, with the largest source of uncertainty being the inclination of the systems.
In edge-on systems, where the inclination correction is minimal, the HI velocity widths are consistent with the galaxies residing in typical dwarf galaxy-mass haloes \citep[see Section~\ref{sec:infall_times}]{He_et_al_19}.

Related to their dynamical mass measurements is whether individual UDGs of similar stellar mass obey the scaling between dark matter halo mass and total globular cluster mass or number \citep{Blakeslee_Tonry_and_Metzger_97, Blakeslee_99, Spitler_and_Forbes_09, Harris_Harris_and_Alessi_13, Harris_Blakeslee_and_Harris_17, Burkert_and_Forbes_20}.
UDGs have been found to host a wide range of globular cluster (GC) numbers, including some with significantly more GCs than normal dwarf galaxies of similar luminosity \citep[e.g.][]{Beasley_and_Trujillo_16, van_Dokkum_et_al_17, Amorisco_et_al_18, Lim_et_al_18, Lim_et_al_20, Forbes_et_al_20, Danieli_et_al_22}. 
Such rich GC systems might indicate the galaxies formed in `overmassive' dark matter haloes \citep{Beasley_et_al_16, Toloba_et_al_23, Forbes_and_Gannon_24}, or, equivalently, formed less stellar mass than expected for their halo mass \citep[often referred to as the `failed galaxy' scenario, e.g.][]{van_Dokkum_et_al_15a, Forbes_et_al_20}.
Generally, reconciling dynamical mass measurements with the GC number-halo mass relation requires such galaxies to reside in cored dark matter haloes \citep{Gannon_et_al_22, Gannon_et_al_23}.
The formation of dark matter cores due to outflows driven by stellar feedback \citep[e.g.][]{Navarro_Eke_and_Frenk_96, Read_and_Gilmore_05, Governato_et_al_10, Maccio_et_al_12, Pontzen_and_Governato_12} may also be a process that results in the formation of UDGs \citep[e.g.][]{Di_Cintio_et_al_17, Chan_et_al_18, Carleton_et_al_19, Martin_et_al_19, Trujillo-Gomez_et_al_21, Trujillo-Gomez_et_al_22}.

Recently, \citet{Gannon_et_al_22} found that UDGs in the Perseus cluster with few GCs have lower stellar velocity dispersions (lower dynamical masses) than those with rich GC systems.
This would be expected if they follow the $N_\mathrm{GC}$-$M_{200}$ relation, and would imply haloes that have efficiently formed stars (i.e.\ low $M_{200}$ for a fixed $M_\ast$) have either less efficiently formed GCs, or preferentially disrupted them.
Alternatively, the radial profile of haloes could differ between GC-rich and GC-poor galaxies \citep[e.g.\ if GC-rich galaxies are formed in high-concentration haloes,][]{Trujillo-Gomez_et_al_22}.
There are also outliers such as NGC1052-DF2 and NGC1052-DF4, which have low dynamical masses but high GC luminosity fractions \citep{van_Dokkum_et_al_18b, van_Dokkum_et_al_19_DF4}.
Such dark matter-deficient galaxies could perhaps form due to tidal stripping of their dark matter haloes by nearby massive galaxies \citep{Safarzadeh_and_Scannapieco_17, Ogiya_18, Jing_et_al_19, Doppel_et_al_21, Jackson_et_al_21a, Maccio_et_al_21, Moreno_et_al_22, Ogiya_et_al_22}, but may require alternative formation scenarios (e.g.\ tidal dwarf galaxies, \citealt{Bournaud_et_al_07, Lelli_et_al_15, Roman_et_al_21, Poulain_et_al_22}; dwarf galaxy collisions, \citealt{Silk_19, Shin_et_al_20, Lee_et_al_21, van_Dokkum_et_al_22}; expansion due to GC feedback, \citealt{Trujillo-Gomez_et_al_22}).
Explanations of dynamical differences between GC-rich and GC-poor galaxies must therefore also take into account the environment in which the galaxies reside, given GC-rich UDGs are often found in galaxy clusters, while isolated field UDGs have very few GCs \citep{Jones_et_al_23}.

In this work, we use simulations of galaxies from the E-MOSAICS project \citep[MOdelling Star cluster population Assembly In Cosmological Simulations within EAGLE,][]{P18,K19a} to explore alternative explanations for the relationship between inferred dynamical mass and GC numbers.
First, we consider whether differing formation histories of galaxies also result in differing dynamical masses.
Cluster galaxies may have truncated formation histories due to early infall times and subsequent quenching.
Therefore, the mass profiles of their dark matter haloes may be more related to those at higher redshifts, which have lower concentrations but are more compact due to their smaller virial radii \citep[e.g.][]{Bullock_et_al_01}.
If GC richness correlates with infall redshift \citep[e.g.][]{Mistani_et_al_16, Carleton_et_al_21}, then a correlation between GC richness and dynamical mass may be expected.
Any of these scenarios may be affected by the tidal stripping of dark matter haloes within group/cluster environments, thus this effect must also be taken into account.

We also consider whether the dynamics of GC-rich and GC-poor galaxies are systematically different, leading to differing inferred dynamical masses.
We suggest that differing formation histories of galaxies may lead to a correlation between GC richness and the amount of dispersion support in the host galaxy.
Galaxies with low velocity dispersions (typically GC-poor galaxies) could be oblate discs observed at low inclinations, such that mainly the vertical motions in the disc contribute to the stellar velocity dispersions, while GC-rich galaxies are largely dispersion supported.

This paper is ordered as follows.
In Section~\ref{sec:methods} we briefly describe the E-MOSAICS simulations and galaxy selection.
Section~\ref{sec:results} presents the main results of this work on correlations between GC system richness and galaxy enclosed masses/kinematics.
Section~\ref{sec:discussion} discusses comparisons of the results of this work with observed galaxies and implications for the formation of UDGs.
Finally, the key results of this work are summarised in Section~\ref{sec:summary}.

\section{Methods}
\label{sec:methods}

In this section, we briefly describe the galaxy and GC formation simulation used in this work (Section~\ref{sec:emosaics}), selection of galaxies and GCs from the simulation (Section~\ref{sec:selection}) and spurious numerical effects that impact analysis of the simulation (Section~\ref{sec:heating}).

\subsection{E-MOSAICS simulation}
\label{sec:emosaics}

The E-MOSAICS project \citep{P18, K19a} is a suite of cosmological hydrodynamical simulations, based on the EAGLE (Evolution and Assembly of GaLaxies and their Environments) galaxy formation model \citep{S15, C15}, which incorporates subgrid models for the formation and evolution of stellar clusters \citep[MOSAICS,][]{Kruijssen_et_al_11, P18}.
Here we only briefly describe the EAGLE and E-MOSAICS models, and refer interested readers to the above works for full details.

EAGLE is a suite of cosmological hydrodynamical simulations of galaxy formation and evolution with a $\Lambda$ cold dark matter cosmogony \citep{S15,C15} whose data have been released to the community \citep{McAlpine_et_al_16}.
The simulations adopt cosmological parameters that are consistent with those inferred by the \citet{Planck_2014_paperXVI}, namely $\Omega_\mathrm{m} = 0.307$, $\Omega_\mathrm{\Lambda} = 0.693$, $\Omega_\mathrm{b} = 0.04825$, $h = 0.6777$ and $\sigma_8 = 0.8288$.
The simulations are performed with a highly modified version of the $N$-body Tree-PM smoothed particle hydrodynamics code \gadget\ \citep[last described by][]{Springel_05}.
The EAGLE model includes subgrid routines describing element-by-element radiative cooling \citep{Wiersma_Schaye_and_Smith_09}, pressure-dependent star formation that reproduces the observed \citet{Kennicutt_98} star-formation law \citep{Schaye_and_Dalla_Vecchia_08}, stellar evolution and stellar mass loss \citep{Wiersma_et_al_09}, the growth of supermassive black holes through gas accretion and mergers \citep{Rosas-Guevara_et_al_15} and feedback associated with both star formation and black hole growth \citep{Booth_and_Schaye_09}.
The simulations lack the resolution and physics to model the cold, dense phase of the interstellar medium.
Therefore, to prevent artificial fragmentation, cold and dense gas is not allowed to cool below temperatures corresponding to an equation of state $P_\mathrm{eos} \propto \rho^{4/3}$, normalised to $T_\mathrm{eos} = 8000 \K$ at $n_\mathrm{H} = 10^{-1} \cmcubed$.
To ensure the emergence of a realistic galaxy population, the parameters governing star formation and black hole feedback were calibrated to reproduce the present-day galaxy stellar mass function, the sizes of disc galaxies and black hole-stellar mass relation \citep{C15}.
The EAGLE simulations have been shown to broadly reproduce a wide variety of observables, including the Tully-Fisher relation and passive galaxy fractions \citep{S15}, the evolution of the galaxy stellar mass function \citep{Furlong_et_al_15} and galaxy sizes \citep[for normal galaxies with $M_\ast \gtrsim 10^{9.5} \Msun$,][]{Furlong_et_al_17}, cold gas properties \citep{Lagos_et_al_15, Crain_et_al_17}, galaxy star formation rates and colours \citep{Furlong_et_al_15, Trayford_et_al_17} and galaxy morphologies \citep{Bignone_et_al_20, Pfeffer_et_al_23a}.

Galaxies (subhaloes) are identified in the simulation using the two-part method described by \citet{S15}.
First, dark matter structures are identified using the friends-of-friends (FOF) algorithm \citep{Davis_et_al_85}.
Next, bound substructures (subhaloes/galaxies) are then identified using the \subfind\ algorithm \citep{Springel_et_al_01, Dolag_et_al_09}.
The galaxy in each FOF group containing the particle with the lowest gravitational potential is considered to be the \textit{central} galaxy, while all others are considered to be \textit{satellite} galaxies.
Galaxy merger trees were constructed from the subhalo catalogues using the D-TREES algorithm \citep{Jiang_et_al_14, Qu_et_al_17}.
We use the merger trees to determine the infall redshift for each satellite galaxy.
We define infall redshift as the first snapshot redshift, after reaching peak subhalo mass as a central, at which a galaxy becomes a satellite.
The requirement for reaching peak mass excludes cases where a galaxy may only briefly be considered a satellite during early major mergers.

The MOSAICS star cluster model \citep{Kruijssen_et_al_11, P18} is coupled to the EAGLE model by treating star clusters as subgrid components of baryonic particles.
Each star particle may host its own subgrid population of star clusters, which is created at the time of star formation (i.e.\ when a gas particle is converted into a star particle).
The star clusters then form and evolve according to local properties within the simulation (i.e.\ local gas and dynamical properties) and adopt the properties of the host particle (i.e.\ positions, velocities, ages, and abundances).
Star cluster formation is determined by two properties: the cluster formation efficiency \citep[CFE, i.e.\ the fraction of stars formed in bound clusters,][]{Bastian_08} and the shape of the initial cluster mass function (a power law or a \citealt{Schechter_76} function with a high-mass exponential truncation, $\Mcstar$).
In the fiducial E-MOSAICS model, the CFE is determined by the \citet{Kruijssen_12} model (where CFE scales with local gas pressure), while $\Mcstar$ is determined by the \citet{Reina-Campos_and_Kruijssen_17} model (where $\Mcstar$ increases with local gas pressure, except in regions limited by high Coriolis or centrifugal forces).
Star cluster formation within each particle is treated stochastically, such that the subgrid clusters may be more massive than the host particle and the mass is conserved only for an ensemble of star particles \citep[for details, see][]{P18}.
After formation, star clusters lose mass by stellar evolution (following the EAGLE model), two-body relaxation that depends on the local tidal field strength \citep{Lamers_et_al_05b, Kruijssen_et_al_11} and tidal shocks from rapidly changing tidal fields \citep{Gnedin_Hernquist_and_Ostriker_99, Prieto_and_Gnedin_08, Kruijssen_et_al_11}.
The complete disruption of clusters by dynamical friction (i.e.\ assuming they merge to the centre of their host galaxy) is treated in post-processing at every snapshot \citep{P18}.

This work discusses galaxies from the E-MOSAICS simulation of a periodic cube with side-length $L = 34.4$~comoving~Mpc \citep{Bastian_et_al_20}.
The simulation initially has $2 \times 1034^3$ particles, using an equal number of baryonic and dark matter particles, with dark matter particle masses of $1.21 \times 10^6 \Msun$ and initial gas particle masses of $2.26 \times 10^5 \Msun$.
The simulation uses a gravitational softening length fixed in comoving units ($1.33$~comoving~kpc) until $z=2.8$, and in proper units ($0.35 \kpc$) thereafter.
The simulation was performed using the `recalibrated' EAGLE model \citep[see][]{S15}.
Though four star cluster formation models are simulated in parallel \citep[setting both the CFE and $\Mcstar$ to be constant or environmentally varying, see][]{Bastian_et_al_20}, this work focuses on the fiducial E-MOSAICS model (with environmentally-varying star cluster formation).
The fiducial E-MOSAICS model produces star cluster populations which are broadly consistent with many observed relations, including the `blue tilt' of GC colour distributions \citep{Usher_et_al_18}, the age-metallicity relations of GC systems \citep{K19b, Kruijssen_et_al_20, Horta_et_al_21a}, the radial distributions of GC systems \citep{K19a, Reina-Campos_et_al_22a}, the scaling relations of young star clusters \citep{Pfeffer_et_al_19b}, the fraction of stars contained in GCs \citep{Bastian_et_al_20}, the UV luminosity function of high redshift proto-globular clusters \citep{Pfeffer_et_al_19a, Bouwens_et_al_21}, the high-mass truncation of GC mass functions \citep{Hughes_et_al_22} and the metallicity distributions of GC systems \citep{Pfeffer_et_al_23b}.
However, as discussed in detail by \citet{P18} and \citet{K19a}, the simulations over-predict the number of low-mass and high-metallicity GCs, which is likely a consequence of insufficient disruption of GCs by tidal shocks due to an overly smooth interstellar medium in the simulations \citep[EAGLE does not model the cold, dense interstellar medium phase, see][]{S15}.

\subsection{Galaxy and GC selection}
\label{sec:selection}

We select dwarf galaxies with stellar masses in the range $2\times 10^8 < M_\ast / \mathrm{M}_{\sun} < 6 \times 10^8$, giving $501$ galaxies at $z=0$ ($188$ are satellites galaxies, $73$ are satellite galaxies in haloes with $M_{200} > 10^{13} \Msun$).
The mass range is chosen to be similar to the stellar masses of galaxies in \citet{Gannon_et_al_22}.
At the resolution of the simulation (and factoring in stellar-evolutionary mass loss for the particles), the mass limits select galaxies with $\approx 1500$-$5000$ stellar particles.
Such galaxies are typically formed in dark matter haloes with masses $\approx 10^{10.7} \Msun$ in the simulation.
We do not select galaxies by size or surface brightness, given that dwarf galaxies in EAGLE simulations are already generally too large and have typical sizes comparable with UDGs \citep[projected half-mass radii $R_{50} \approx 2 \kpc$,][]{S15}.
This is a result of numerical limitations in the simulations, i.e.\ the temperature floor for dense gas, which artificially thickens discs \citep{Benitez-Llambay_et_al_18}, and spurious heating from interactions of stellar and dark matter particles \citep[see also following section]{Ludlow_et_al_19}.

E-MOSAICS models GC formation assuming both young and old star clusters form via a common mechanism.
Therefore, to compare populations of GCs, we select star clusters with masses $> 10^{4.5} \Msun$ and ages $> 2 \Gyr$ at $z=0$ (i.e.\ clusters that have already undergone significant mass loss through stellar evolution).
All such star clusters that are bound to a galaxy (subhalo) according to \subfind\ are taken to be part of the galaxy's GC population.

\subsection{Galaxy dynamics and spurious heating}
\label{sec:heating}

We elect not to directly investigate the stellar kinematics of dwarf galaxies in the simulation, as they are affected by the spurious heating of stellar orbits by interactions with dark matter particles.
This process will occur in any simulation when dark matter particles are significantly more massive than baryonic particles \citep{Ludlow_et_al_19, Ludlow_et_al_21, Ludlow_et_al_23, Wilkinson_et_al_23}.
Galaxies affected by spurious heating become larger and rounder, and the stellar dynamics becomes more dispersion supported.
By contrast, the effect on the dark matter profiles is significantly smaller, with a slight increase in halo concentration within the stellar half-mass radius due to the mass segregation of the more massive dark matter particles \citep{Ludlow_et_al_19}.
Older galaxies will be more affected by spurious heating than younger galaxies, given the longer timescale for heating to occur.
For the particle masses of the EAGLE high-resolution model, this particularly affects galaxies in haloes with masses $M_\mathrm{200} \lesssim 10^{12} \Msun$ \citep{Ludlow_et_al_21}.

Therefore, we instead investigate the dynamics of `star-forming' gas in the simulation, for which the limiting factor is the radiative cooling model.
EAGLE does not model the cold gas phase, and instead has a polytropic equation of state $P_\mathrm{eos} \propto \rho^{4/3}$ with a temperature floor at $8 \times 10^3 \K$ \citep{S15}.
The cooling floor implies a velocity dispersion floor that increases weakly with density/pressure ($\sigma \approx [13, 17, 23] \kms$ at $P/k = [10^3, 10^4, 10^5] \K \cmcubed$).
This is slightly higher than the stellar velocity dispersion found for NGC1052-DF2 and DF4 \citep{Danieli_et_al_19, Emsellem_et_al_19, Shen_et_al_23}, but is generally smaller than the velocity dispersions of normal dwarf galaxies with stellar masses between $10^8$ and $10^9 \Msun$, which have stellar velocity dispersions in the range $20$ to $100 \kms$ \citep[e.g.][]{Norris_et_al_14}.

\section{Results}
\label{sec:results}

\subsection{Enclosed dark matter masses and GC system richness}
\label{sec:SM-Mdm}

\begin{figure*}
    \includegraphics[width=\textwidth]{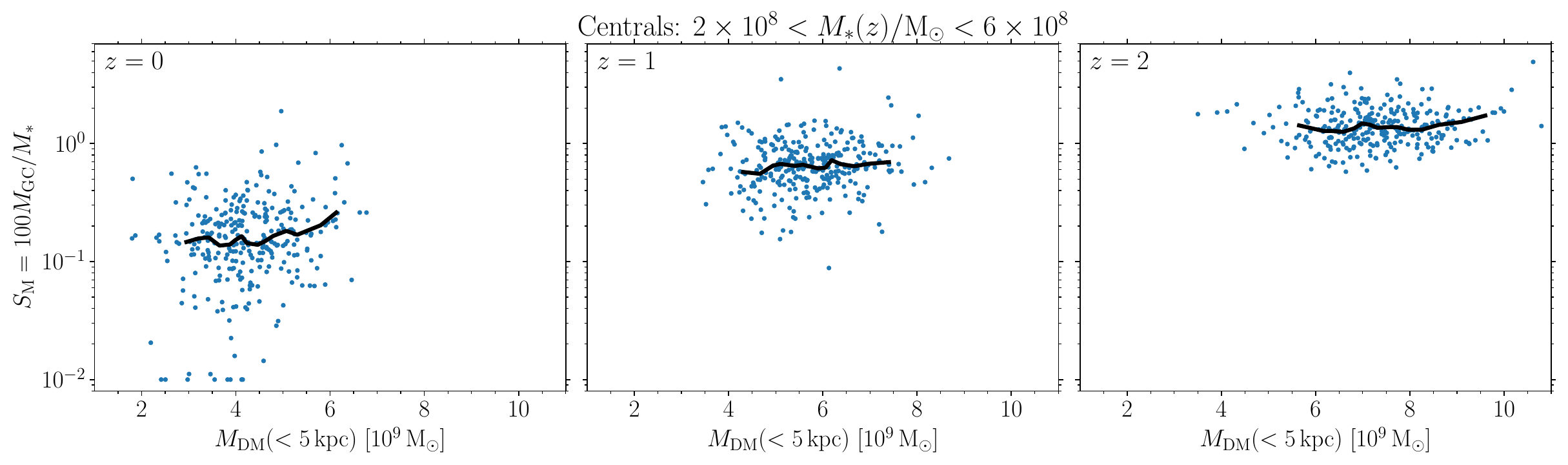}
    \caption{GC specific masses ($S_M$) compared with enclosed dark matter mass within $5 \kpc$ for central dwarf galaxies in the E-MOSAICS simulation. The panels show galaxies with stellar masses $2 \times 10^8 < M_\ast(z) / \Msun < 6 \times 10^8$ at different redshifts ($z=0,1,2$ in the left, middle and right panels, respectively). Galaxies with specific masses below the plotted range are shown at $S_M = 1 \times 10^{-2}$ (all of these galaxies have $S_M = 0$). Note that the axis limits are identical in each panel so they may be directly compared. Only weak correlations are found between $S_M$ and $M_\mathrm{DM}(<5\kpc)$ at all redshifts (Spearman correlation coefficients $\approx 0.1$, with $p$-values $\approx 0.1$), indicating that correlations between enclosed mass within $5 \kpc$ and GC richness are not expected in the E-MOSAICS model.}
    \label{fig:SM-Mdm}
\end{figure*}

In Figure~\ref{fig:SM-Mdm} we first compare the relationship between GC specific mass (the ratio of total GC mass and galaxy stellar mass, $S_M = 100 M_\mathrm{GC} / M_\ast$) and dark matter mass ($M_\mathrm{DM}$) within $5 \kpc$ for central dwarf galaxies (i.e.\ excluding satellites of larger haloes) at different redshifts.
Such a correlation might be expected if $S_M$ depends on the halo concentration\footnote{For example, relative to a typical NFW halo concentration for our galaxies of $c_\mathrm{NFW} = 10$ and assuming fixed $M_{200}$ and $r_{200}$, the enclosed dark matter mass within $5 \kpc$ would decrease (increase) by a factor of two for $c_\mathrm{NFW} = 5$ ($c_\mathrm{NFW} = 20$). For an NFW halo this relative difference becomes larger at smaller radii.} or $M_\ast / M_{200}$.
The radius limit of $5 \kpc$ was chosen to be similar to the dynamical mass measurements of UDGs \citep[e.g.][]{Gannon_et_al_22}, however the correlations for $3 \kpc$ and $10 \kpc$ radii are nearly identical to the $5 \kpc$ limit.
The comparison is limited to only central galaxies so that effects from the tidal stripping of dark matter haloes are excluded.
Galaxies are selected within the same stellar mass range ($2 \times 10^8 < M_\ast(z) / \Msun < 6 \times 10^8$) at each redshift (i.e. the figure does not show an evolutionary sequence).
At all redshifts ($z=0$, 1, 2), there are only weak correlations between $S_M$ and $M_\mathrm{DM}(<5 \kpc)$, having Spearman correlation coefficients $\approx 0.1$ (with $p$-values $\approx 0.1$).
Therefore, for dwarf galaxies in the E-MOSAICS simulations, differences in $S_M$ at fixed $M_\ast$ are not driven by differences in dark matter halo mass.

However, the typical values for both $S_M$ and $M_\mathrm{DM}(<5\kpc)$ increase with redshift, with $S_M$ increasing by a factor $\sim 10$ from $z=0$ to $z=2$, and $M_\mathrm{DM}(<5\kpc)$ increasing by a factor $\sim 2$.
For the enclosed dark matter mass, this difference is because, at fixed halo mass, halo virial radii are smaller at earlier times, even though haloes are less concentrated ($r_{200}$ decreases from $\approx 85$ to $40 \kpc$ from $z=0$ to $z=2$, while $c_\mathrm{NFW}$ decreases from $\approx 10$ to $6$).
For the stellar mass range considered here, the total halo mass for galaxies at different redshifts are roughly constant ($M_\mathrm{200} \approx 7 \times 10^{10} \Msun$) over the entire redshift range.
The high $S_M$ at earlier times is driven by the higher CFE due to higher natal gas pressure in the E-MOSAICS model \citep{P18}, as well as the young ages of clusters at high redshift which are yet to undergo significant mass loss.
In this stellar mass range, we expect at least a $0.5$~dex decrease in $S_M$ from initial to $z=0$ values \citep{Bastian_et_al_20}.
Thus if the formation of a galaxy/halo was truncated at early times (e.g.\ via infall into a proto-cluster), the galaxy might plausibly have both elevated $S_M$ and enclosed dark matter mass.
In such a case, the galaxy would appear to reside in an `overmassive' halo based on its enclosed mass, despite forming in a dwarf galaxy-mass halo ($M_\mathrm{200} \approx 7 \times 10^{10} \Msun$).
However, as we discuss in the next section, tidal stripping of dark matter haloes once they become satellites needs to be taken into account.

\subsection{Enclosed masses and cluster infall times}
\label{sec:infall_times}

\begin{figure*}
    \includegraphics[width=0.495\textwidth]{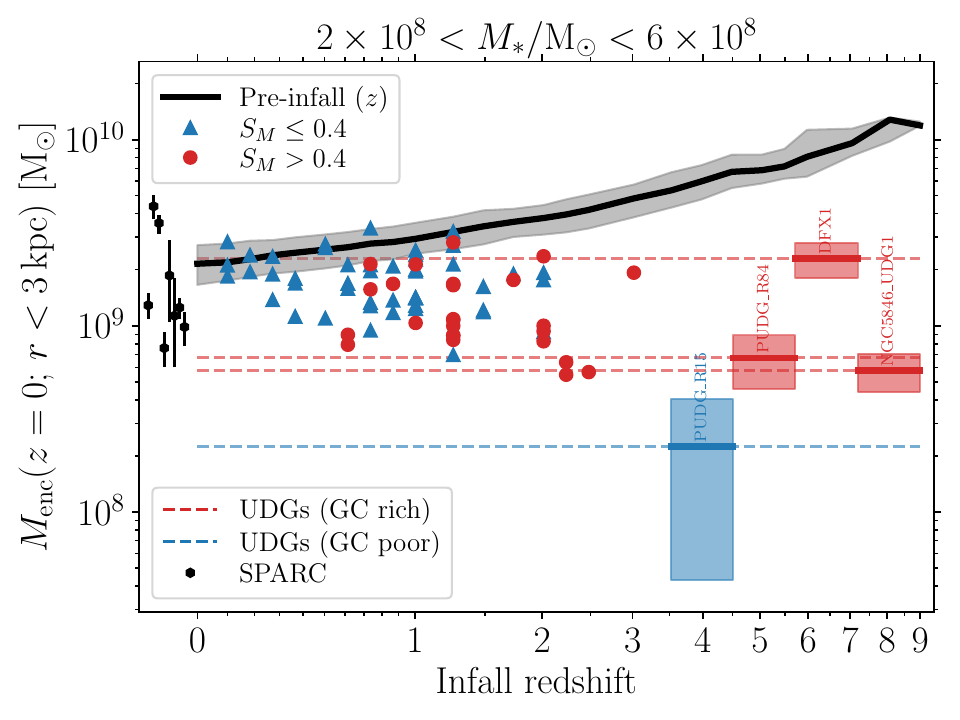}
    \includegraphics[width=0.495\textwidth]{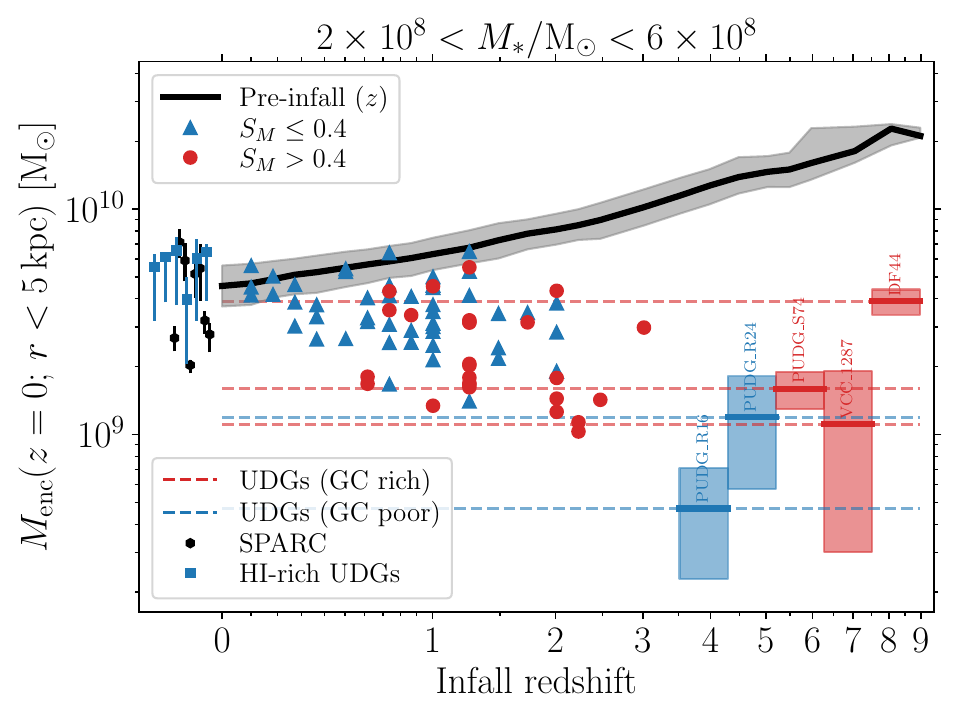}
    \caption{Enclosed masses within $3\kpc$ (left panel) and $5\kpc$ (right panel) at $z=0$ as a function of infall redshift for E-MOSAICS dwarf galaxies in galaxy groups and clusters (FOF group mass $10^{13} < M_{200} \leq 10^{13.71} \Msun$). Red circles indicate GC-rich galaxies ($S_M > 0.4$) while blue triangles show GC-poor galaxies ($S_M < 0.4$). The median ``pre-infall'' masses at each redshift (i.e.\ for central galaxies) are shown by the solid black line, with the grey shaded region showing the 16\ts{th}--84\ts{th} percentiles. Early infalling galaxies (which are typically GC rich) have lost more dark matter mass to tidal stripping than late infalling galaxies (which are typically GC poor). For comparison, the dashed lines (mean mass, shown the full redshift range to indicate unknown infall redshifts) and shaded regions (indicating $1\sigma$ uncertainties) show the dynamical masses from stellar velocity dispersions for GC-rich (red) and GC-poor (blue) UDGs from \citet{Gannon_et_al_22}. The dynamical masses of GC-rich UDGs agree well with the masses of simulated GC-rich galaxies. The dynamical masses of GC-poor UDGs are generally significantly lower than those of simulated GC-poor galaxies. The blue squares \citep[given HI-rich UDGs are GC poor,][]{Jones_et_al_23} show the dynamical masses of edge-on HI-rich field UDGs \citep{He_et_al_19}. Black hexagons show the dynamical masses of normal HI-rich, high inclination ($i > 60$ degrees) dwarf galaxies from SPARC \citep{Lelli_et_al_16}. Both samples of HI rich galaxies (the edge-on field UDGs and SPARC dwarfs) reasonably agree with the masses of $z\approx 0$ simulated galaxies, though with larger scatter in the observed samples.}
    \label{fig:Mdyn}
\end{figure*}

The sample of UDGs with dynamical masses in \citet{Gannon_et_al_22} are from galaxy groups and clusters.
Therefore for comparison we select E-MOSAICS dwarf galaxies which are satellites at $z=0$ in the seven most massive FOF groups with $10^{13} < M_{200} \leq 10^{13.71} \Msun$.
The most massive FOF group is limited by the simulation volume ($34.4^3 \Mpc^3$), and therefore the simulation does not contain galaxy clusters as massive as the Perseus cluster \citep[$M_{200} \approx 10^{15} \Msun$,][]{Aguerri_et_al_20}.
The main impact is the volume is missing dwarf galaxies with infall redshifts $>3$, as higher mass haloes may have earlier infalling satellite galaxies (e.g. $z_\mathrm{infall} \lesssim 2$ for $M_{200} \lesssim 10^{12} \Msun$ and $z_\mathrm{infall} \lesssim 3$ for $M_{200} \lesssim 10^{13.7} \Msun$ for the dwarf galaxy mass range we consider, which would imply a maximum $z_\mathrm{infall} \approx 4$ for $M_{200} = 10^{15} \Msun$ if the scaling continues to higher masses).

In Figure~\ref{fig:Mdyn} we compare the total enclosed masses (i.e. sum of the masses of all gas, dark matter, stellar and black hole particles) within $3$ and $5 \kpc$ at $z=0$ for the E-MOSAICS dwarf galaxies as a function of the infall redshift (the redshift at which the galaxies become satellites in a larger halo).
The radius limits are chosen to be similar to the effective radii range ($2.7$ to $5.2 \kpc$) for UDGs in the sample of \citet{Gannon_et_al_22}.
The galaxies are divided into ``GC-rich'' (red circles) and ``GC-poor'' (blue triangles) populations, with the division at a specific mass of $S_M = 0.4$ (twice the typical value of $S_M$ at $z=0$, Figure~\ref{fig:SM-Mdm}; which corresponds to the $20$ GC limit used by \citealt{Gannon_et_al_22} at $M_\ast = 5 \times 10^8 \Msun$, assuming a typical GC mass of $10^5 \Msun$ for dwarf galaxies, \citealt{Jordan_et_al_07_XII}).
GC-rich galaxies typically have earlier infall redshifts ($z_\mathrm{infall} \gtrsim 1$) than GC-poor galaxies, as expected from the increase in $S_M$ with redshift \citep[Figure~\ref{fig:SM-Mdm}; see also][]{Mistani_et_al_16, P18, Carleton_et_al_21}.

For comparison, the black line in Figure~\ref{fig:Mdyn} shows the pre-infall relation, i.e.\ the typical enclosed masses for central galaxies within the same stellar mass range at each snapshot.
Although the enclosed masses for central galaxies increase with redshift (owing to more compact haloes, as discussed in Section~\ref{sec:SM-Mdm}), on average the enclosed masses of satellite galaxies decrease with increasing infall redshift.
This is due to tidal stripping of the dark matter haloes over time, which, interestingly, occurs at a faster rate than tidal stripping of the stellar component (e.g.\ for a galaxy which has lost half its stellar mass within $5 \kpc$, it will have lost $\approx 85$ per cent of its dark matter mass within the same radius).
The faster stripping of dark matter may be due to the radially-biased orbits of dark matter haloes \citep[e.g.][]{Cole_and_Lacey_96, Colin_Klypin_and_Kravtsov_00}.
The impact of gas stripping on the enclosed masses is minor, as it typically only contributes $\approx 10$ per cent of the pre-infall enclosed mass.
This ratio is similar to that found for the SPARC galaxies \citep{Lelli_et_al_16} shown in Figure~\ref{fig:Mdyn} (see below).
Gas stripping is more important for earlier infalling galaxies (approximately 40 per cent of the dwarf galaxies with $z_\mathrm{infall} \leq 1$ retain gas at $z=0$, compared to $\approx 3$ per cent with $z_\mathrm{infall} > 1$).
In general, GC-rich simulated galaxies tend to have lower enclosed masses than GC-poor galaxies due to their earlier infall times (i.e.\ more time available for stripping of the dark matter haloes).
Galaxies with infall redshifts $\lesssim 0.2$ have enclosed masses similar to central (isolated) galaxies at $z = 0$.
Such late-infalling galaxies typically have group-/cluster-centric distances $\gtrsim R_{200}$\footnote{Though beyond the group `virial radius' the galaxies at $r_\mathrm{group} > R_{200}$ are bound to the group. This simply indicates that a 1D radius may not describe well the triaxial shapes of massive haloes, such as recently merged groups or where the most bound/central galaxy is offset from the centre of mass.}.

In Figure~\ref{fig:Mdyn} we also compare the inferred dynamical masses of UDGs from the sample compiled by \citet{Gannon_et_al_22}.
The sample is limited to galaxies with dynamical masses from stellar kinematics.
For the UDG sample, the effective radii are in the range $2.7$ to $5.2 \kpc$, and we assign the galaxies to the panel with the closest measurement radius ($3$ or $5 \kpc$).
For observed galaxies the infall redshift is unknown, and thus the median measurement is indicated over the full redshift range.
Overall, the enclosed masses from GC-rich simulated galaxies reasonably match the range from GC-rich UDGs.
Thus, based on both their GC numbers and dynamical masses, GC-rich UDGs are consistent with being early-forming galaxies which have had much of their dark matter haloes stripped within the clusters.
Most GC-rich UDGs in \citet[NGC 5846 UDG1, PUDG R84, PUDG S74, VCC 1287]{Gannon_et_al_22}, as well as one GC-poor UDG (PUDG R24), have dynamical masses consistent with simulated galaxies with infall redshifts $z \approx 1$-$3$.
However, \citet{Gannon_et_al_22} noted that PUDG R24 has a disturbed morphology and is significantly bluer than other Perseus UDGs, which may indicate recent infall and quenching in the cluster.
The GC-rich UDGs with the highest masses (DFX1, DF44) have dynamical masses consistent with simulated galaxies over the full redshift range.

In contrast, the inferred masses of observed GC-poor UDGs are generally inconsistent with those of the GC-poor simulated galaxies, being a factor of 5--10 lower than expected from the simulations.
Of the GC-poor UDGs, only PUDG R24 is consistent with the lowest enclosed masses found for simulated galaxies.
Overall, the simulations predict the opposite trend to that found for observed UDGs, with GC-poor simulated galaxies typical having \textit{higher} masses than GC-rich galaxies.
For comparison with the $z \approx 0$ simulated galaxies, we include HI-rich dwarf galaxies ($2\times10^8 < M_\ast / \Msun < 6 \times 10^8$) with high inclinations ($i > 60$) from SPARC \citep[\textit{Spitzer} Photometry and Accurate Rotation Curves,][]{Lelli_et_al_16}, where dynamical masses were calculated from the HI rotation curves.
The median dynamical masses from SPARC reasonably agree with the simulated galaxies, though with slightly larger scatter.
In the right panel of Figure~\ref{fig:Mdyn} we also show dynamical masses for edge-on, HI-rich UDGs with $M_\ast > 10^8 \Msun$ from \citet{He_et_al_19}.
Although other HI-rich UDGs also have measured HI rotational velocities, uncertainties in their inclination make their dynamical mass estimates highly uncertain \citep{Karunakaran_et_al_20}.
We convert HI velocity widths to rotation velocities assuming that the galaxies are edge-on, then convert rotation velocities to dynamical masses assuming the peak rotation velocities are reached by $5 \kpc$\footnote{For dwarf galaxies in the SPARC sample with $10^8 < M_\ast / \Msun < 10^9$, around two-thirds of galaxies reach the flat part of the rotation curve by $5 \kpc$ \citep{Lelli_et_al_16}.}.
In the uncertainties for the dynamical masses we include an error on the inclination of 10 degrees and a decrease in the velocity widths accounting for if the rotation velocity at $5 \kpc$ is only 80 per cent of the peak velocity \citep[e.g.\ as for AGC219533,][]{Leisman_et_al_17}.
The GC-poor simulated galaxies with late infall times agree well with HI-rich field galaxies and UDGs, which host few GCs \citep{Jones_et_al_23} and have dynamical masses expected for normal dwarf galaxies at $z=0$.

\begin{figure*}
    \includegraphics[width=0.495\textwidth]{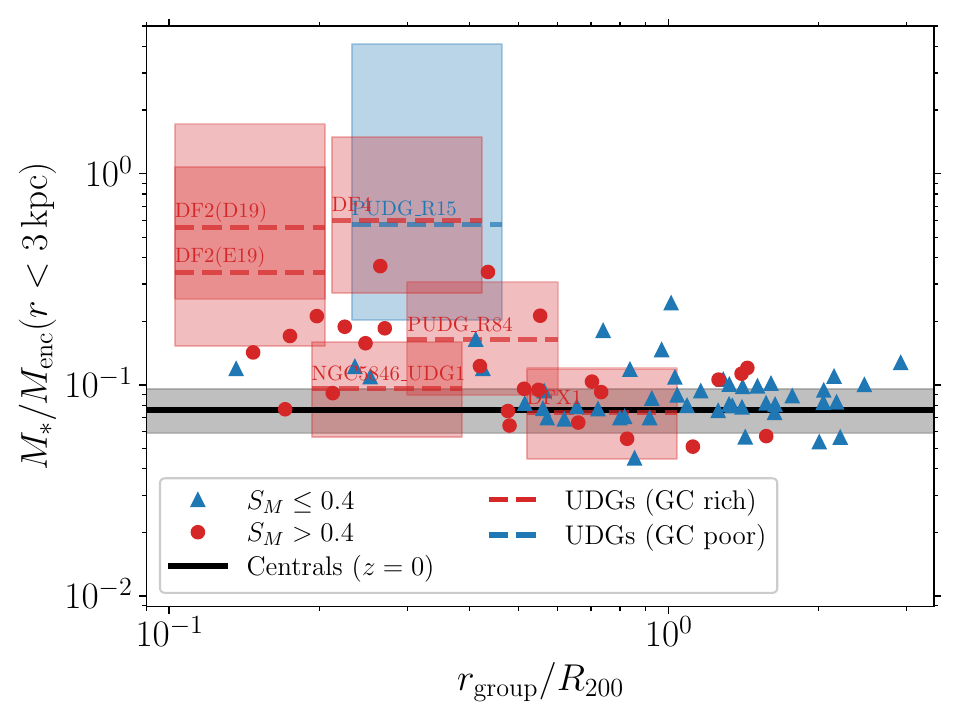}
    \includegraphics[width=0.495\textwidth]{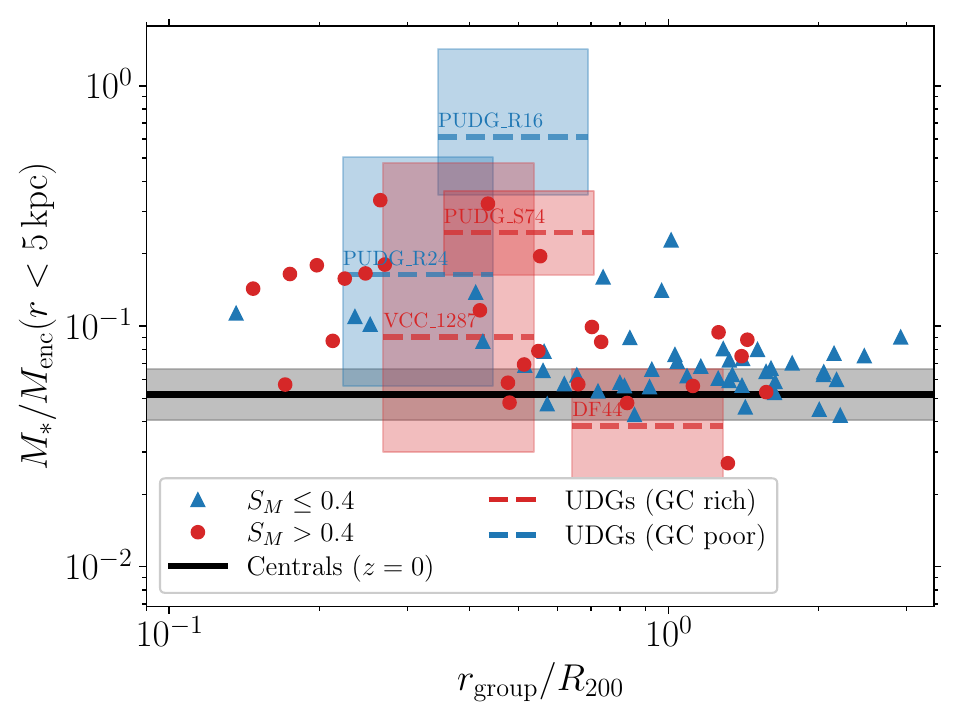}
    \caption{Stellar-to-enclosed mass ratios within $3$ (left panel) and $5 \kpc$ (right panel) as a function of group-/cluster-centric radius (distance to central galaxy of the FOF group, scaled by the group virial radius $R_{200}$) for the same galaxies in Figure~\ref{fig:Mdyn}. Red circles indicate GC-rich galaxies ($S_M > 0.4$) while blue triangles show GC-poor galaxies ($S_M < 0.4$). For reference, the median mass ratios for central galaxies at $z=0$ are shown by the solid black line, with the grey shaded region showing the 16\ts{th}--84\ts{th} percentiles, although we note that the mass ratio evolves with redshift (lower $M_\ast / M_\mathrm{dyn}$ at higher redshifts) as can be inferred from the higher pre-infall (central galaxy) dynamical masses in Figure~\ref{fig:Mdyn}. The dashed lines (mean mass ratio) and shaded regions (indicating $1\sigma$ uncertainties) show the stellar-to-dynamical mass ratios for GC-rich (red) and GC-poor (blue) UDGs from \citet{Gannon_et_al_22}, as well as NGC1052-DF2 (showing two mass estimates from velocity dispersion measurements from \citealt{Danieli_et_al_19} and \citealt{Emsellem_et_al_19}) and NGC1052-DF4 \citep{Shen_et_al_23}. Group-/cluster-centric distances for observed galaxies are assumed to be uncertain by a factor of two, with the minimum distance taken to be the projected distance from the central galaxy in the group/cluster (thus realistically, the distances for observed galaxies are only lower limits). The stellar mass-to-light ratios for observed galaxies are assumed to have an uncertainty of $\pm 0.5 \MLsun$.}
    \label{fig:Mratio}
\end{figure*}

As an alternative way to view the results in Figure~\ref{fig:Mdyn} (and a way to factor out the stellar mass differences between galaxies), in Figure~\ref{fig:Mratio} we instead compare the enclosed stellar-to-enclosed mass ratios ($M_\ast / M_\mathrm{dyn}$) within $3$ and $5\kpc$ as a function of group-/cluster-centric radius.
Comparing stellar-to-enclosed mass ratios also helps to factor out differences in total mass as a function of stellar mass.
For the observed galaxies, the enclosed stellar mass is taken to be half the total stellar mass, given the dynamical masses are estimated within the half-light radius \citep{Gannon_et_al_22}.
The trends in Figure~\ref{fig:Mratio} follow the trends in Figure~\ref{fig:Mdyn} for the simulated galaxies.
Galaxies at smaller group-centric distances (earlier infall times) tend to be GC rich and have elevated $M_\ast / M_\mathrm{dyn}$ to central galaxies at $z=0$.
Galaxies at larger group-centric distances (later infall times) tend to be GC poor and have $M_\ast / M_\mathrm{dyn}$ that is similar to central galaxies at $z=0$.
For the galaxies with the highest stellar-to-enclosed mass ratios ($\gtrsim 0.2$), the mass ratios are not sensitive to the exact radius used, and the results are also similar for mass ratios within $10 \kpc$. 
The UDGs from \citet{Gannon_et_al_22} also generally follow the simulation trends, with those at the largest distances (DFX1, DF44) having the lowest $M_\ast / M_\mathrm{dyn}$.
Again, the GC-poor Perseus UDGs R15 and R16 are the largest outliers, having higher than expected $M_\ast / M_\mathrm{dyn}$ compared to the simulated GC-poor galaxies.

In the left panel of Figure~\ref{fig:Mratio} we also show the dark matter-deficient UDGs NGC1052-DF2 (using the stellar velocity dispersions from both \citealt{Danieli_et_al_19} and \citealt{Emsellem_et_al_19}) and NGC1052-DF4 \citep{Shen_et_al_23}.
Dynamical masses were estimated using the same method as \citet{Gannon_et_al_22}, i.e.\ using the mass estimator for dispersion-supported galaxies from \citet{Wolf_et_al_10}.
Given their projected distance to the elliptical galaxy NGC1052, we assume that both are members of the galaxy group, although note that tip of the red giant branch distance measurements may place one galaxy, or both, outside the group due to the relative distance of $\approx 2 \Mpc$ between the UDGs \citep{Shen_et_al_21}.
Both galaxies are taken to be GC rich given their high luminosity fraction in GCs \citep[3-4 per cent, ][]{van_Dokkum_et_al_18b, van_Dokkum_et_al_19_DF4}.
The stellar-to-dynamical mass ratios for both NGC1052-DF2 and NGC1052-DF4 ($\approx 0.6$) are consistent with the highest mass ratios found in the simulations ($\approx 0.4$) within their uncertainties.
In this case, the dark matter-deficient galaxies are consistent with tidal stripping of the dark matter halo \citep[c.f.][]{Safarzadeh_and_Scannapieco_17, Ogiya_18, Jing_et_al_19, Doppel_et_al_21, Jackson_et_al_21a, Maccio_et_al_21, Moreno_et_al_22, Ogiya_et_al_22}.
Both NGC1052-DF2 and NGC1052-DF4 have evidence of tidal distortions which would support such a scenario \citep{Montes_et_al_20, Keim_et_al_22}.
If the UDGs reside outside the NGC1052 group \citep[see][]{Shen_et_al_21} then alternative formation scenarios are needed, such as galaxy collisions \citep{Silk_19, Shin_et_al_20, Lee_et_al_21, van_Dokkum_et_al_22} or expansion through feedback from star cluster formation \citep{Trujillo-Gomez_et_al_22}.

Interestingly, the Perseus cluster GC-poor UDGs R15 and R16 have similar $M_\ast / M_\mathrm{dyn}$ to NGC1052-DF2 and NGC1052-DF4, as well as relatively small cluster-centric radii in projection ($r_\mathrm{clust} / R_{200} \approx 0.2$--$0.3$; which were selected to be matched in projected cluster-centric radii to other Perseus GC-rich UDGs, see \citealt{Gannon_et_al_22}).
This could indicate both galaxies are also affected by significant tidal stripping of their dark matter haloes.
However, the simulations show this process is more likely to affect GC-rich galaxies, given their earlier infall times than GC-poor galaxies.

To summarise the results of this section, overall, the simulations agree well with the dynamical masses of both field galaxies and GC-rich group/cluster galaxies.
In the latter case, this also suggests agreement in mass loss from tidal stripping of galaxies and their dark matter haloes in groups/clusters.
Therefore, disagreement in the enclosed and dynamical masses of GC-poor group/cluster galaxies between the simulations and observations likely has a different origin, which we explore in the following sections.

\subsection{Dynamics of star-forming systems}
\label{sec:dynamics}

\begin{figure}
    \includegraphics[width=\columnwidth]{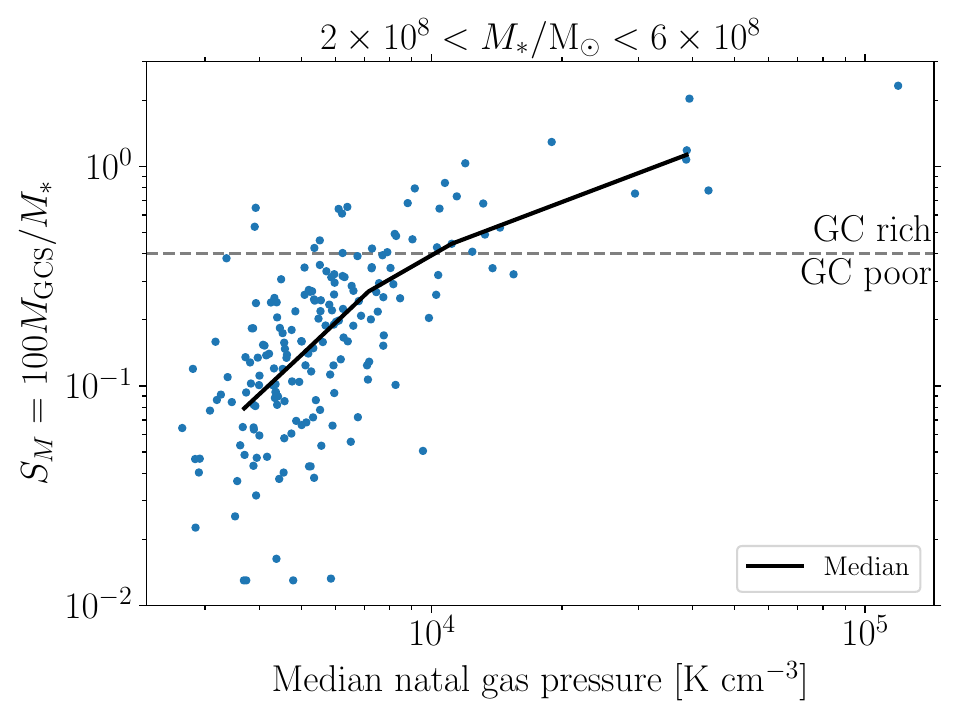}
    \caption{GC specific masses ($S_M$) versus median natal gas pressure of stellar particles for E-MOSAICS satellite dwarf galaxies. The strong correlation between $S_M$ and natal pressure is driven by the scaling of cluster formation efficiency with natal pressure in the E-MOSAICS model \citep{P18}. The transition from GC-poor ($S_M < 0.4$) to GC-rich ($S_M > 0.4$) galaxies occurs at $P/k \approx 10^4 \K \cmcubed$.}
    \label{fig:SM-P}
\end{figure}

Given that tidal stripping of dark matter haloes may not explain the dynamical masses of GC-poor UDGs in clusters, we now consider an alternative explanation: GC-poor dwarf galaxies are not dispersion-supported systems.
In this case, the stellar velocity dispersions of observed GC-poor galaxies may underestimate the actual mass, depending on the inclination of the system.
We suggest that galactic conditions leading to high-pressure star formation, and thus efficient GC formation, also lead to dispersion-supported stellar systems (spheroids).
Conversely, conditions favouring low-pressure star formation (inefficient GC formation) instead lead to rotationally-supported discs.
This may explain why early-type galaxies have richer GC systems than late-type galaxies \citep{Georgiev_et_al_10}.

In the E-MOSAICS model, the main difference between GC-rich and GC-poor dwarf galaxies is the natal gas pressures of star formation.
We demonstrate this in Figure~\ref{fig:SM-P}, showing the relationship between $S_M$ and median natal gas pressure in satellite dwarf galaxies.
In the simulations, this relationship is due to the CFE scaling directly with the natal gas pressure \citep[see][]{P18}.
At low median pressures ($P/k \sim 10^{3.5} \K \cmcubed$), there is significant scatter in $S_M$ at a given median pressure due to factors such as temporal and spatial variations in pressure and stochasticity in the GC formation model.
Typically, galaxies become GC-rich ($S_M > 0.4$) at a median $P/k \gtrsim 10^4 \K \cmcubed$.

\begin{figure}
    \includegraphics[width=\columnwidth]{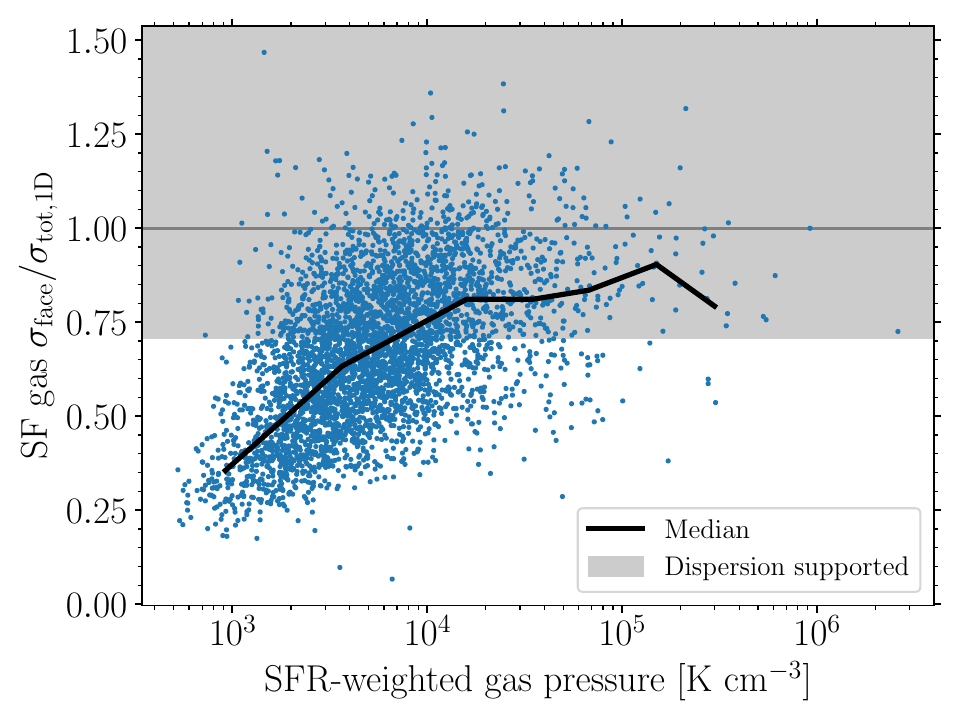}
    \caption{Correlation between dispersion support of star-forming gas and the gas pressure (weighted by the star formation rate of gas particles) for the progenitors of galaxies with stellar masses $2 \times 10^8 < M_\ast / \Msun < 6 \times 10^8$ at $z=0$. Only galaxies with at least 500 star-forming gas particles are shown. The dispersion support of the star-forming gas is measured as the ratio between the face-on 1D velocity dispersion, $\sigma_\mathrm{face}$, and the total 1D velocity dispersion, $\sigma_\mathrm{tot,1D}$, after orienting the systems by the spin of the star-forming gas. The dark grey line at $\sigma_\mathrm{face} / \sigma_\mathrm{tot,1D} = 1$ indicates the velocity dispersion ratio of an isotropic spheroid. The shaded region with $\sigma_\mathrm{face} / \sigma_\mathrm{tot,1D} > 1/\sqrt{2}$ indicates where the ratio of kinetic energies is at least 50 per cent, as an approximate estimate for the onset of dispersion-supported dynamics. The dispersion support of star-forming systems correlates with pressure, such that at low gas pressures ($P/k \sim 10^3 \K \cmcubed$) they are rotationally supported, while at higher pressures ($P/k \gtrsim 10^4 \K \cmcubed$) they are largely dispersion supported.}
    \label{fig:sigmaRatio}
\end{figure}

Next, we turn to the connection between the gas pressure and dispersion-supported kinematics.
To demonstrate this, in Figure~\ref{fig:sigmaRatio} we compare the dynamics of star-forming gas in the progenitors of galaxies with $2 \times 10^8 < M_\ast / \Msun < 6 \times 10^8$ at $z=0$ (recalling that we compare gas kinematics, rather than stellar kinematics, due to the effect of spurious heating on stellar orbits, Section~\ref{sec:heating}).
Only central galaxies (at any redshift) are included to avoid effects such as tidal disruption of satellites.
The figure shows the ratio of the face-on 1D velocity dispersion ($\sigma_\mathrm{face}$, i.e. the vertical velocity dispersion of the disc in the case of an oblate rotator) to the total 1D velocity dispersion ($\sigma_\mathrm{tot,1D} = \sigma_\mathrm{tot,3D} / \sqrt{3}$) as a function of star formation rate-weighted gas pressure.
We investigate $\sigma_\mathrm{face} / \sigma_\mathrm{tot,1D}$, rather than other measures of rotation support (e.g.\ $v/\sigma$), in order to understand how the line-of-sight velocity dispersion (i.e.\ that used to determine dynamical mass) may be affected in rotating systems.
The star formation rate weighting accounts for the likelihood of gas particles being converted to stars depending on pressure in the EAGLE model \citep{S15}.
To orient the systems face on, the spin of the star-forming gas system was also determined by weighting the gas particles by their star formation rate.
Only galaxies with $\geq 500$ star-forming gas particles are included in the figure so that the dynamics and spin of the gas are reasonably resolved.
At a given pressure there is relatively large scatter in $\sigma_\mathrm{face} / \sigma_\mathrm{tot,1D}$, which may be driven by processes such as accretion, mergers and stellar feedback altering the dynamics of the systems.

Figure~\ref{fig:sigmaRatio} shows that at higher pressures ($P/k \gtrsim 10^4 \K \cmcubed$) the star-forming gas systems become mostly dispersion supported ($\sigma_\mathrm{face} / \sigma_\mathrm{tot,1D} > 1/\sqrt{2}$).
At lower pressures, $\sigma_\mathrm{face} / \sigma_\mathrm{tot,1D}$ decreases with pressure, and the star-forming gas becomes increasingly rotationally supported.\footnote{Note that there is an implied floor in $\sigma_\mathrm{face} / \sigma_\mathrm{tot,1D}$. The pressure floor in EAGLE (see Section~\ref{sec:heating}) implies a minimum $\sigma_\mathrm{face} = 13 \kms$, though in practice we often find velocity dispersions below this limit. The galaxies in Figure~\ref{fig:sigmaRatio} have a maximum $\sigma_\mathrm{tot,1D} \approx 50 \kms$, leading to the minimum ratio $\sigma_\mathrm{face} / \sigma_\mathrm{tot,1D} \approx 0.25$.}
Given the expected trend between natal pressure and $S_M$, galaxies formed with low natal gas pressures are expected to be rotationally-supported systems and host few GCs.

\subsection{Origin of gas pressure--kinematics correlation}
\label{sec:origin}

\begin{figure}
    \includegraphics[width=\columnwidth]{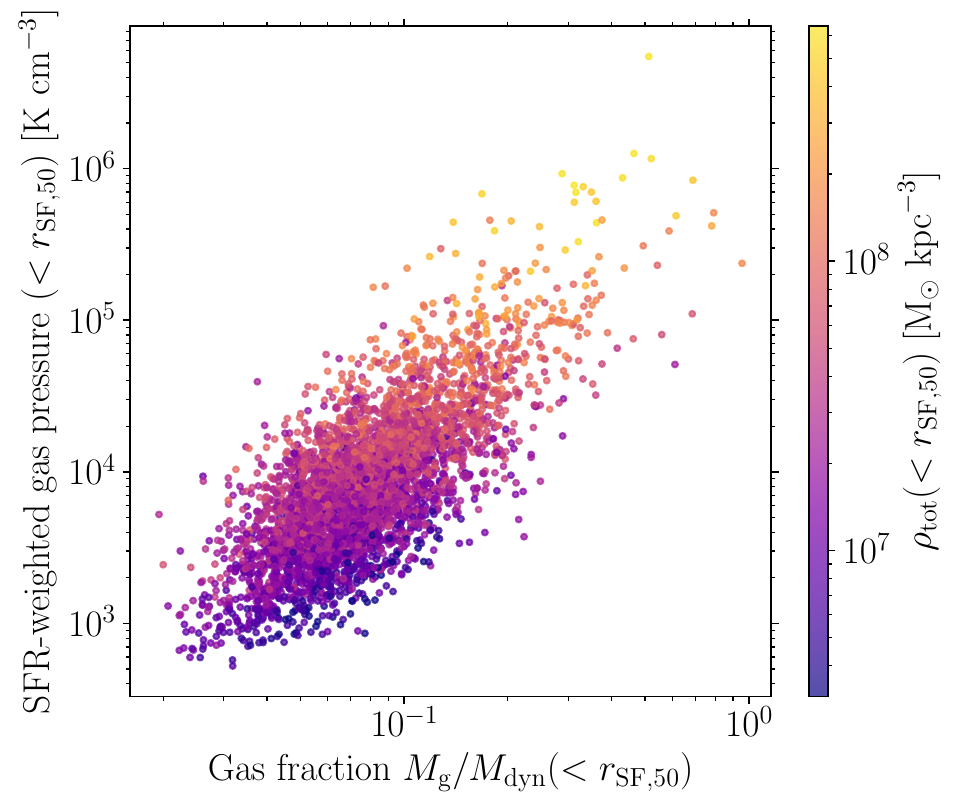}
    \caption{Correlation between gas pressure (weighted by the star formation rate of the gas particles) and gas fraction for the progenitors of galaxies with stellar masses $2 \times 10^8 < M_\ast / \Msun < 6 \times 10^8$ at $z=0$ (as in Figure~\ref{fig:sigmaRatio}). The points for each galaxy are coloured by the total mass density. The galaxy properties are calculated within the radius containing 50 per cent of star formation ($r_\mathrm{SF,50}$) to focus on the main regions of star formation. Higher gas fractions and mass densities both lead to higher pressures for star formation.}
    \label{fig:fgas}
\end{figure}

We now investigate the underlying cause of the relation between gas pressure and dynamics.
The rotational support of gas in a galaxy is expected to scale with the gas fraction of the galaxy, $v_\mathrm{rot}/\sigma \propto 1 / f_\mathrm{gas}$ \citep[e.g.][]{Dekel_Sari_and_Ceverino_09, Genzel_et_al_11}.
In fact, this has been found previously for simulations of UDG formation, where dispersion-supported galaxies tend to accrete gas earlier, and thus achieve higher gas fractions, than rotation-supported galaxies \citep{Cardona-Barrero_et_al_20}.
Similarly, since gas pressure is expected to scale with gas surface density \citep[as $P \propto {\Sigma_\mathrm{g}}^2$, under the assumption of hydrostatic equilibrium; e.g.][]{Schaye_01a, Blitz_and_Rosolowsky_04, Krumholz_and_McKee_05}, we may also expect a correlation between gas pressure and gas fraction in the galaxy.

Figure~\ref{fig:fgas} shows how the gas pressure scales with the gas fraction for the progenitors of galaxies with $2 \times 10^8 < M_\ast/ \Msun < 6 \times 10^8$ at $z=0$.
Only central galaxies with at least 500 star-forming gas particles are included in the figure for consistency with Figure~\ref{fig:sigmaRatio}.
The gas fractions and SFR-weighted pressure are calculated within the radius containing 50 per cent of star formation ($r_\mathrm{SF,50}$) to focus on the main region of star formation in the galaxies, though similar trends are obtained for other choices in radii (e.g.\ radii containing 90 per cent of star formation).
There is a clear trend of increasing gas pressure with increasing gas fraction in galaxies, though with some scatter at a given gas fraction.
However, the contribution of the total mass (rather than just gas mass) to the gas pressure must also be taken into account.
In the figure the points for each galaxy are coloured by the total mass density within $r_\mathrm{SF,50}$.
At a given gas fraction, higher mass densities result in higher gas pressures, which drives scatter in the gas fraction-gas pressure correlation.
This can be expected to drive scatter in the correlations between gas pressure and kinematics (i.e.\ Figure~\ref{fig:sigmaRatio}), as well as GC abundance and kinematics.
Overall, we arrive at a coherent picture where the gas fractions in dwarf galaxies as they are forming can result in a correlation between the dynamics of galaxies and their GC system richness.

Of course, the kinematics of galaxies can also be influenced by other processes, such as galaxy mergers or harassment \citep{Moore_et_al_96, Moore_et_al_99}. Such processes are expected to result in dynamical heating of the galaxies, which could alter any initial correlation between kinematics and GC system richness.
To determine how much effect major mergers may have on the dynamics of galaxies, we search the galaxy merger trees for significant mergers (those capable of drastically changing the dynamics of the system, i.e.\ major mergers) which occur late in the formation of the galaxies (such that a new disc dominating the mass of the system is unlikely to form)\footnote{The discussion here differs from cases where mergers alter the gas dynamics, and subsequent star formation, in the system \citep[e.g. mergers leading to UDG formation through increased spin and size of gas discs,][]{Di_Cintio_et_al_19, Wright_et_al_21}. In that case, the kinematics and GC richness of the system would largely be set by the subsequent star formation (assuming it dominates the total stellar mass). We focus here on cases where the majority of star formation occurs prior to the merger.}.
For all simulated galaxies in the stellar mass range $2 \times 10^8 < M_\ast / \Msun < 6 \times 10^8$, we find that only $\approx$5 per cent of galaxies have major mergers (stellar mass ratio $m_2/m_1> 0.25$) where the stellar mass of the descendant of the merger is at least 50 per cent of the $z=0$ stellar mass of the galaxy.
Thus, in most cases, late galaxy mergers are unlikely to have played a significant role in setting the stellar dynamics of dwarf galaxies.
For GC-rich ($S_M > 0.4$) dwarf galaxies $\approx 12$ per cent of galaxies have undergone late major mergers, compared to $\approx 4$ per cent for GC-poor ($S_M \leq 0.4$) dwarf galaxies.
Thus late major mergers are more important in setting the dynamics of GC-rich galaxies, but still in a minority of systems.

In clusters, GC-rich galaxies, which are expected to have earlier infall times (Section~\ref{sec:infall_times}), might be subjected to galaxy harassment over a longer period than GC-poor galaxies.
However, we are currently unable to investigate the importance of this effect given the spurious particle heating in the simulation (see Section~\ref{sec:heating}).

\section{Discussion}
\label{sec:discussion}

\subsection{Comparison with observations}
\label{sec:comparison}

As we found in Sections~\ref{sec:dynamics} and \ref{sec:origin}, galaxies that host few GCs may be rotationally-supported systems as conditions favouring disc formation lead to low-pressure star formation (inefficient GC formation).
This is consistent with observations showing that early-type dwarf galaxies have richer GC systems than late-type dwarf galaxies \citep{Georgiev_et_al_10}.
Therefore, in rotationally-supported galaxies observed at low inclinations, the measured velocity dispersion is only representative of the vertical support in the disc, and may significantly underestimate the actual mass of the system (e.g.\ by a factor $16$ for $\sigma_\mathrm{face} / \sigma_\mathrm{tot,1D} = 0.25$, given $M_\mathrm{dyn} \propto \sigma^2$).
If the GC-poor UDGs in Figure~\ref{fig:Mdyn} \citep[from][]{Gannon_et_al_22} also have $\sigma_\mathrm{face} / \sigma_\mathrm{tot,1D} \approx 0.25$--$0.5$, then their dynamical masses would be 4--16 times larger, which would bring them in line with the dynamical masses of both the GC-rich UDGs and predictions from the simulations.
Clearly, direct observation of rotation or its absence in such galaxies is needed to test the scenario.

To date, only two UDGs have spatially-resolved stellar kinematic profiles: NGC1052-DF2 \citep{Emsellem_et_al_19} and DF44 \citep{van_Dokkum_et_al_19}.
Both galaxies have rich GC systems \citep[$\sim$3-4 per cent of the galaxy luminosity,][]{van_Dokkum_et_al_18b, van_Dokkum_et_al_19} and thus, based on the results of this work, we would not expect them to be rotationally-supported galaxies.
NGC1052-DF2 appears to have prolate rotation, which may be evidence of a merger\footnote{It is possible that prolate rotation could also occur due to high-speed galaxy collisions, though in the simulation of \citet{Lee_et_al_21} the resulting galaxy is an oblate rotator.}, tidal stirring or stripping \citep{Emsellem_et_al_19}, rather than oblate rotation that would be expected if it were a disc galaxy\footnote{Interestingly, the GC system of NGC1052-DF2 may have rotation which is perpendicular to the stellar rotation \citep{Lewis_et_al_20}. However, in this work, we concentrate on the stellar rotation.}.
DF44 shows no evidence for rotation \citep[$V/\sigma < 0.12$][]{van_Dokkum_et_al_19}, but has an inferred dynamical mass ($M_\mathrm{dyn} \sim 4 \times 10^9 \Msun$) that is consistent with the simulated galaxies (Figure~\ref{fig:Mdyn}).
On the other hand, HI-rich field UDGs \citep[which have few GCs,][]{Jones_et_al_23} show clear rotation in HI \citep{Leisman_et_al_17, Sengupta_et_al_19, Gault_et_al_21, Shi_et_al_21}.

Although no GC-poor UDGs in clusters currently have resolved kinematic profiles, oblate rotating galaxies may be detected statistically by comparing their shapes, as long as the number of galaxies is large enough to sample a range of inclinations.
In fact, \citet{Lim_et_al_18} found that UDGs in the Coma cluster with lower GC specific frequencies ($S_N$) have lower axis ratios ($b/a$), in agreement with our scenario.
Similarly, nucleated dwarf galaxies also have rounder shapes than non-nucleated dwarf galaxies \citep{Ferguson_and_Sandage_89, Ryden_and_Terndrup_94, Lisker_et_al_07, Venhola_et_al_19, Poulain_et_al_21}.
Such a correlation would also be expected if the formation of nuclear clusters is related to the formation of GCs (e.g.\ \citealt{Sanchez-Janssen_et_al_19}; see \citealt{Neumayer_et_al_20} for a review of the formation mechanisms of nuclear clusters).

\begin{figure*}
    \includegraphics[width=\columnwidth]{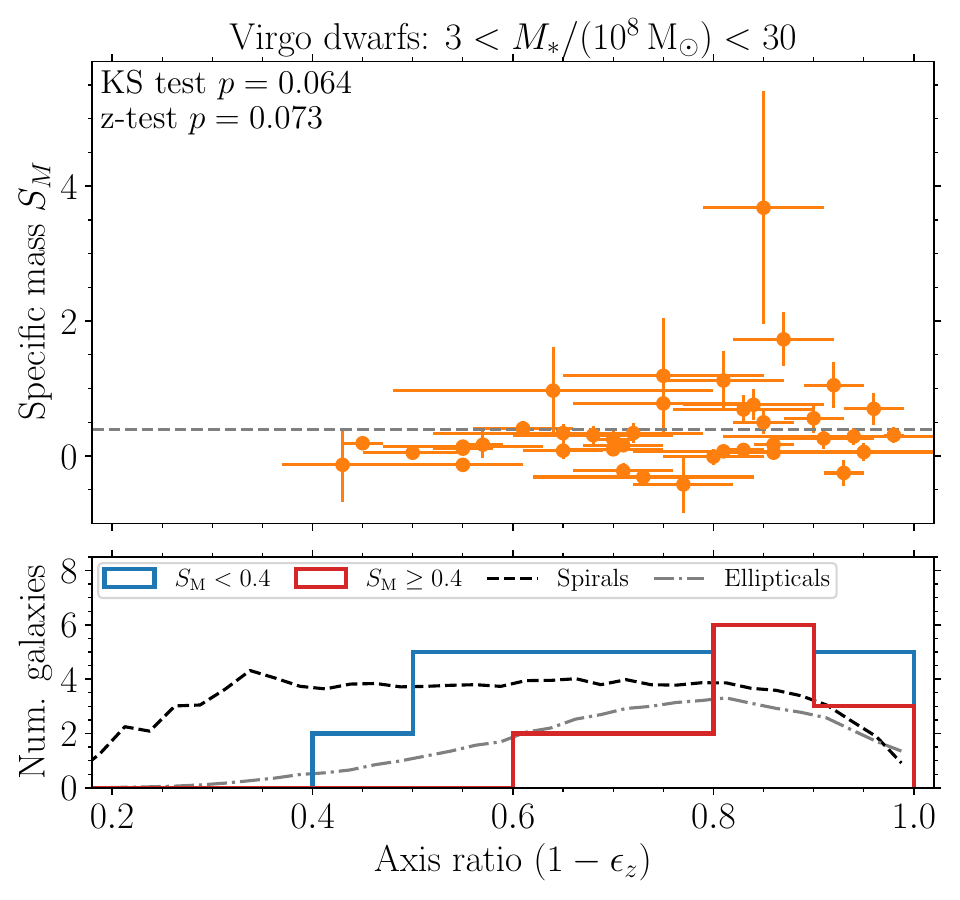}
    \includegraphics[width=\columnwidth]{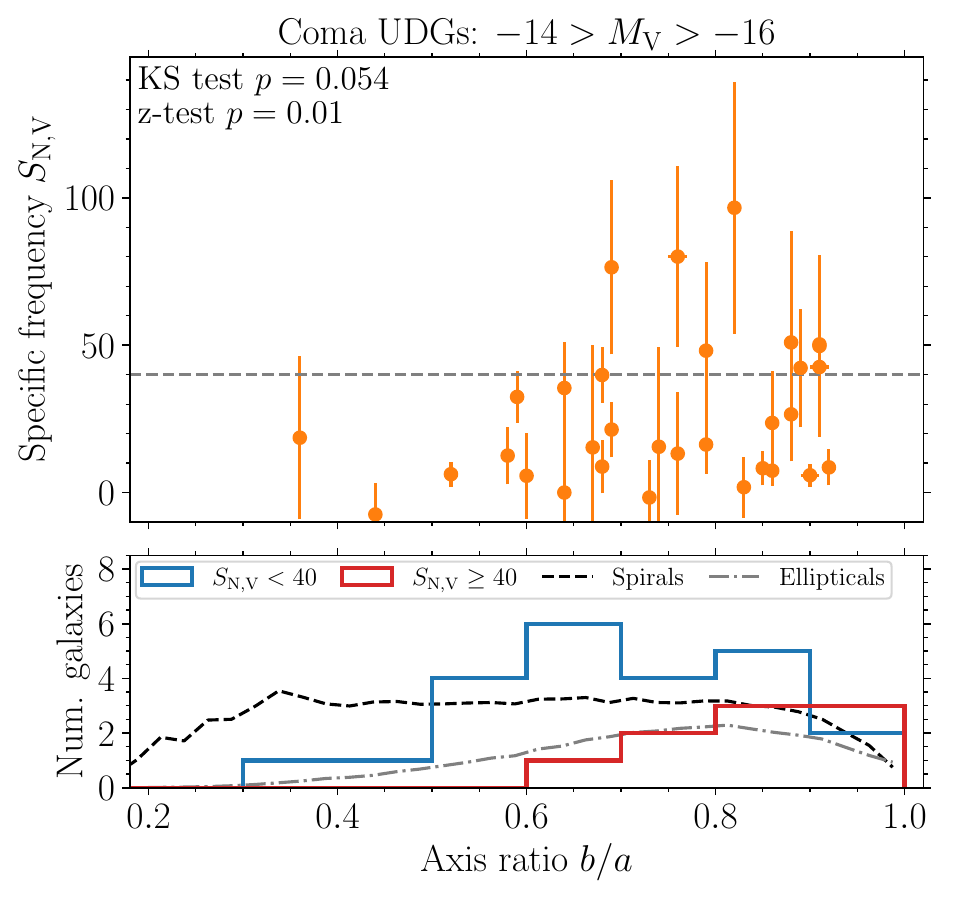}
    \includegraphics[width=\columnwidth]{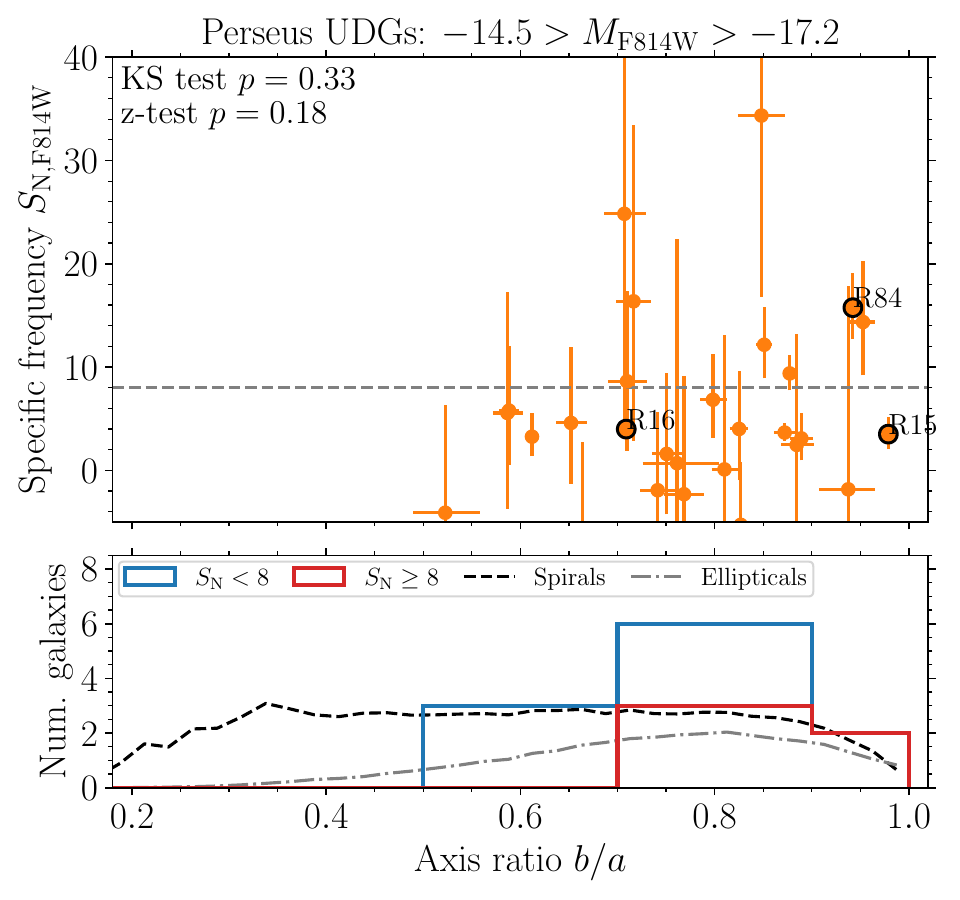}
    \includegraphics[width=\columnwidth]{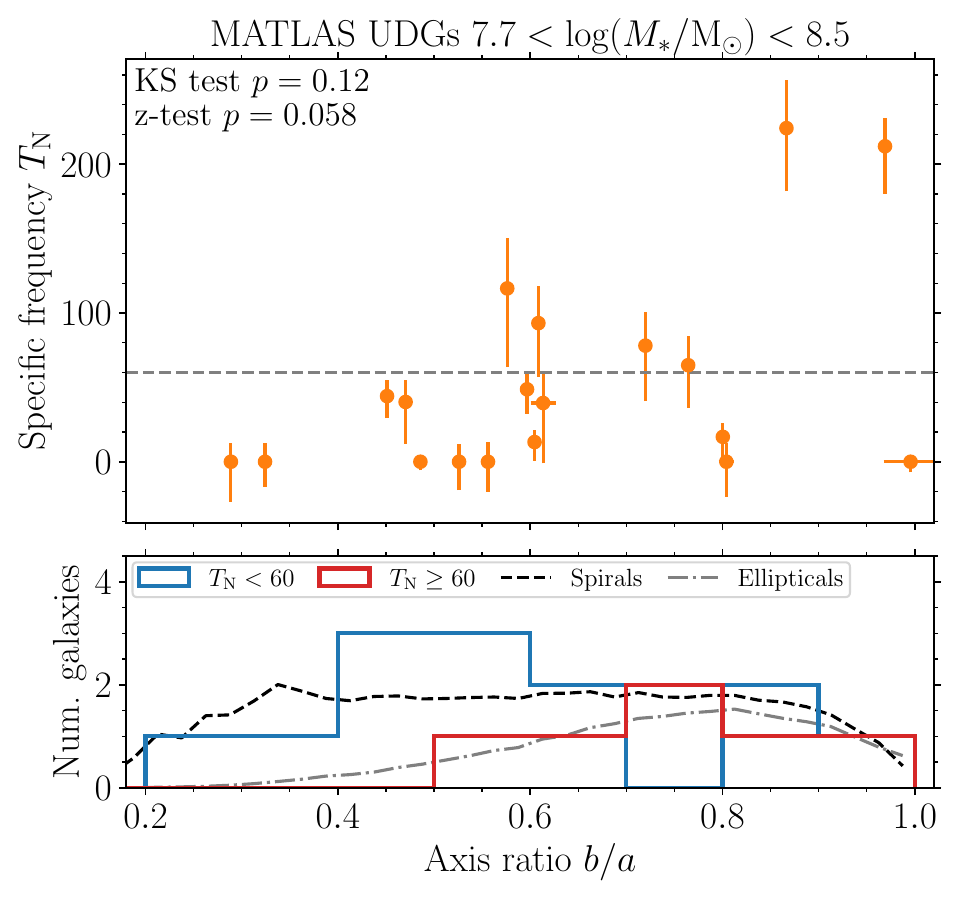}
    \caption{GC system richness (specific frequency or mass) compared with galaxy axis ratios for different observed samples of dwarf galaxies. The top left panels show normal dwarf galaxies from the ACS Virgo Cluster Survey \citep{Ferrarese_et_al_06, Peng_et_al_08}. The top-right panels show UDGs in the Coma cluster \citep{Yagi_et_al_16, Forbes_et_al_20}. The bottom left panels show UDGs in the Perseus cluster (Janssens et al., in prep.) with three UDGs in common with the \citet{Gannon_et_al_22} sample highlighted (PUDG R15, R16, R84). The bottom-right panels show UDGs from the MATLAS survey \citep{Buzzo_et_al_24}. For each galaxy sample, the bottom subpanel shows the axis ratio histograms for `GC poor' and `GC rich' galaxies, where the division is indicated as the grey dashed lines in the main panels at approximately twice the median specific mass/frequency for each sample (top left: $S_M = 0.4$; top right: $S_\mathrm{N,V} = 40$; bottom left: $S_\mathrm{N,F814W} = 8$; bottom right: $T_N = 60$). For reference, also shown in the subpanels are the axis ratio distributions of bright ($M_r < -19.77$) spiral and elliptical galaxies in the Sloan Digital Sky Survey \citep{Rodriguez_and_Padilla_13}, with the spiral and elliptical distributions normalised to the number of GC-poor and GC-rich galaxies, respectively. The titles of the main panels show the mass/luminosity limits used for each galaxy sample. To test if the shape distributions for GC-rich and GC-poor galaxies are statistically different, a Kolmogorov--Smirnov and $z$ test were performed for each galaxy sample, with the resulting $p$ values shown in the main panels. In all samples the GC-rich galaxies are, on average, rounder than the GC-poor galaxies, which may indicate GC-poor galaxies have flatter intrinsic shapes than GC-rich galaxies (though this difference is not statistically significant for Perseus UDGs).}
    \label{fig:obs}
\end{figure*}

We test this further in Figure~\ref{fig:obs} by comparing the GC system richness (specific frequency or specific mass) with stellar axis ratios for a number of observed dwarf galaxy samples.
The top left panel of the figure shows dwarf galaxies from the ACS Virgo Cluster Survey, with GC specific masses from \citet{Peng_et_al_08} and ellipticities from \citet{Ferrarese_et_al_06}.
The top right panel of the figure shows UDGs in the Coma cluster, with specific frequencies ($S_{N,V} = N_\mathrm{GC} 10^{0.4 [M_{V} + 15]}$) from \citet{Forbes_et_al_20} and axis ratios from \citet{Yagi_et_al_16}.
The bottom left panel of the figure shows UDGs in the Perseus cluster, with specific frequencies ($S_{N,\mathrm{F814W}} = N_\mathrm{GC} 10^{0.4 [M_\mathrm{F814W} + 15]}$) and axis ratios from Janssens et al. (in prep.).
The bottom right panel of the figure shows UDGs from the MATLAS survey (mainly field galaxies and galaxy groups), with specific frequencies ($T_{N} = N_\mathrm{GC} / [M_\ast / 10^9 \Msun]$) and axis ratios from \citet{Buzzo_et_al_24}.
The lower mass/luminosity limits for the three UDG samples are chosen to be approximately similar with a limit $M_\ast \gtrsim 5 \times 10^7 \Msun$.
The ACS Virgo Cluster Survey targets slightly higher mass dwarf galaxies, with a lower limit of $M_\ast > 3 \times 10^8 \Msun$.
We also note that specific frequencies calculated in different photometric bands are not directly comparable \citep[e.g.\ there is an average factor of $\approx2.2$ difference between the $V$ and ACS $z$ bands,][]{Peng_et_al_08}.
Therefore, we compare each galaxy sample separately given the different analysis methods used.

For each galaxy sample, the division between `GC-rich' and `GC-poor' galaxies is taken at approximately twice the median specific frequency/mass (as in Section~\ref{sec:infall_times}; grey dashed lines in each panel of the figure, with the exact values listed in the caption).
For both the Virgo and Coma clusters, the differences in axis ratios between GC-rich and GC-poor galaxies are statistically significant ($p < 0.1$) according to both Kolmogorov--Smirnov and $z$ tests.
The Perseus UDGs are not statistically different in either test.
For the MATLAS UDGs the GC-rich and GC-poor galaxies are statistically different according to a $z$ test (i.e.\ statistically different means), but not a Kolmogorov--Smirnov test, which may be a reflection of the smaller sample size.
GC-rich galaxies have, on average, higher axis ratios (rounder shapes) than GC-poor galaxies, with an almost complete lack of GC-rich galaxies at $b/a \lesssim 0.6$.
Instead, GC-poor galaxies are found to have a wide range of axis ratios, which may be consistent with intrinsically flattened galaxies (oblate or slightly triaxial systems) observed at random inclinations \citep[e.g.\ see figure 7 in][]{Sanchez-Janssen_et_al_16}. 
This is found both for UDGs (Coma and MATLAS samples, though not in Perseus) and normal dwarf galaxies in the Virgo cluster.

Comparison of the axis ratio distributions of GC-poor galaxies with spiral galaxies \citep{Rodriguez_and_Padilla_13} shows similarly flat distributions, though the cluster dwarf samples (Virgo, Coma, Perseus) tend to lack very flattened galaxies ($b/a \lesssim 0.3$-$0.4$) compared to spiral galaxies.
This could be a result of formation differences between dwarf galaxies and spirals, galaxy harassment in clusters, or simply selection bias in the samples \citep[e.g.\ edge-on galaxies are less likely to pass the UDG criteria due to their higher surface brightnesses,][]{He_et_al_19, Cardona-Barrero_et_al_20, Van_Nest_et_al_22}.
In contrast, the GC-rich galaxies have axis ratio distributions which are closer to that of elliptical galaxies \citep[projected axis ratios peaking at $b/a \approx 0.8$, consistent with spheroidal systems with intrinsic 3D axis ratios $C/A \approx 0.6$,][]{Rodriguez_and_Padilla_13}.

In principle, a trend of galaxies with richer GC systems having rounder shapes could also arise from external environmental processes, rather than galaxy formation processes.
For example, GC-rich galaxies, which tend to be found at smaller cluster-centric distances (earlier cluster infall times), may have been more affected by galaxy harassment/tidal heating than GC-poor galaxies.
That the trend may also be found for the MATLAS UDG sample (bottom right panel in Figure~\ref{fig:obs}), where the galaxies are found in much lower density environments \citep{Marleau_et_al_21} than the Virgo, Coma and Perseus clusters, suggests that galaxy harassment is not the (sole) origin of the correlation.

Previous work has modelled the projected axis ratios of UDGs to determine their intrinsic shapes.
\citet{Burkert_17} found that UDGs have prolate shapes.
By contrast, both \citet{Rong_et_al_20} and \citet{Kado-Fong_et_al_21} instead found that UDGs have oblate-triaxial shapes.
\citet{Rong_et_al_20} suggest the difference in results arises because they considered triaxial models, while \citet{Burkert_17} considered only purely oblate or prolate models.
Here we instead argue that UDGs are not a homogeneous population, but rather a composite of oblate (perhaps slightly triaxial) rotators and dispersion-supported spheroidal galaxies (i.e.\ the GC-poor and GC-rich galaxies, respectively).
This could bias the determination of their intrinsic shape when modelling the axis ratio distributions.

\citet{Rong_et_al_20} also found that UDGs in clusters tend to have rounder shapes at smaller cluster-centric distances.
This is similarly found for dwarf galaxies in the Fornax cluster \citep{Rong_et_al_19} and by the ACS Virgo Cluster Survey \citep{Ferrarese_et_al_06, Peng_et_al_08}.
The trend between galaxy shape and cluster-centric distance would agree with the trend we find between GC richness and infall redshift (Figure~\ref{fig:Mdyn}; i.e. if GC-rich galaxies are typically rounder and earlier-infalling galaxies typically being located at smaller cluster-centric distances, we return to this point in the context of UDGs in the following section).
However, again a correlation of shape and cluster-centric radius could also be caused by galaxy harassment/tidal heating.
Future simulations, which are not affected by spurious particle heating \citep[see Section~\ref{sec:heating}]{Ludlow_et_al_19}, are needed to disentangle whether harassment or turbulent star formation is more important for setting shapes of cluster dwarf galaxies.

\subsection{Formation of UDGs}
\label{sec:UDG_formation}

Given their similar sizes and surface brightnesses \citep[e.g.][]{Leisman_et_al_17}, it is natural to consider whether HI-rich field UDGs \citep[or slightly brighter, bluer dwarfs which have since faded, e.g.][]{Roman_and_Trujillo_17} might be the progenitors of UDGs in clusters which have since had their gas stripped.
Based solely on GC numbers \citep{Jones_et_al_23}, it is unlikely that present-day HI-rich UDGs could be transformed into GC-rich UDGs.
However, it is natural to consider whether HI-rich field UDGs are the progenitors of GC-poor UDGs in clusters.
If the dynamical masses (calculated assuming dispersion-supported systems) from \citet{Gannon_et_al_22} are correct, to match GC-poor UDGs, the HI-rich UDGs would need to undergo significant dynamical evolution to remove most of their central dark matter mass, and at a much higher level of mass loss than found in the simulated galaxies.
This appears unlikely, as dark matter mass loss would need to disproportionately affect GC-poor UDGs compared to GC-rich UDGs, given that the latter do agree with simulated galaxies (Figure~\ref{fig:Mdyn}).
Instead, the dark matter halos of HI-rich UDGs and GC-poor cluster UDGs would have to be significantly different initially (e.g.\ cored dark matter profiles in only cluster galaxies, though cored haloes are less able to survive tidal disruption, \citealt{Errani_et_al_23}; the dark matter profiles of UDGs is further discussed below).
GC-poor cluster UDGs might instead form from baryon-dominated field UDGs \citep{Mancera_Pina_et_al_19, Mancera_Pina_et_al_20, Mancera_Pina_et_al_22, Kong_et_al_22} which have lost their gas within the cluster \citep{Gannon_et_al_23}.
Such galaxies are suggested to form due to weak stellar feedback \citep{Mancera_Pina_et_al_19, Mancera_Pina_et_al_20}, though this would lead to increased star formation and consumption of gas \citep[e.g.][]{C15}, which appears inconsistent with such gas-rich galaxies.
\citet{Sellwood_and_Sanders_22} also show that such baryon-dominated discs would be unstable, thus requiring an increased dark matter fraction.
Reconciling the rotation curves of baryon-dominated UDGs with normal HI-rich galaxies would require their inclinations are systematically overestimated \citep{Karunakaran_et_al_20, Mancera_Pina_et_al_22, Sellwood_and_Sanders_22}.
A larger sample of dynamical mass measurements of edge-on, HI-rich UDGs \citep{He_et_al_19}, where inclination corrections are small, may help resolve the issue.

Alternatively, as we have discussed in this work, the velocity dispersions of GC-poor cluster UDGs might instead underestimate their mass if they are oblate rotating galaxies observed at low inclinations.
This could be due to selection effects, where only oblate galaxies that are (nearly) face-on may be classified as UDGs \citep{He_et_al_19, Cardona-Barrero_et_al_20, Van_Nest_et_al_22}, which would be particularly important for higher mass UDGs (i.e.\ typical spectroscopic targets with $M_\ast \gtrsim 10^8 \Msun$) due to their higher surface brightnesses.
In this case, HI-rich UDGs (or higher redshift analogues) could be the progenitors of GC-poor cluster UDGs.
GC-rich UDGs may instead differ due to formation biases (early halo and star formation) introduced by early infall and subsequent quenching within groups/clusters.
Such an origin was previously advocated by \citet{Carleton_et_al_21}, who also suggested that the large sizes of cluster UDGs may be due to tidal heating \citep[see also][]{Yozin_and_Bekki_15, Safarzadeh_and_Scannapieco_17, Ogiya_18, Carleton_et_al_19, Sales_et_al_20, Tremmel_et_al_20}.
This might imply that the large sizes of GC-rich and GC-poor UDGs occur (on average) through different processes \citep[echoing previous suggestions of different formation processes for GC-rich and GC-poor UDGs,][]{Forbes_et_al_20}.
For example, early cluster infall for GC-rich galaxies (Section~\ref{sec:infall_times}) may lead to enhanced tidal heating, while for GC-poor and HI-rich UDGs the large sizes could largely be due to internal galaxy formation processes (e.g.\ formation in low concentration or high-spin haloes leading to lower density star formation, \citealt{Amorisco_and_Loeb_16, Leisman_et_al_17, Shi_et_al_21, Benavides_et_al_23}; or increase spin and size of gaseous discs due to mergers, \citealt{Di_Cintio_et_al_19, Wright_et_al_21}).
Indeed, some works find that a combination of formation paths are necessary to reproduce UDG numbers in groups and clusters \citep[e.g.][]{Jiang_et_al_19a, Sales_et_al_20, Benavides_et_al_23}.
Interestingly, based on their phase-space locations, cluster UDGs found in the `ancient infall' region \citep[small cluster-centric distances and low relative velocities, with infall times greater than $6.45 \Gyr$ ago, or $z>0.7$, see][]{Rhee_et_al_17} tend to have lower GC numbers than those in other regions of phase space \citep{Forbes_et_al_23}.
If GC-rich UDGs form through tidal heating, this appears at odds with the expectation that tidal heating would be more important for earlier-infalling galaxies, though we note that GC specific frequency of UDGs does decrease with increasing cluster-centric radius in the Coma cluster \citep{Lim_et_al_18}.
Further study is clearly needed, but it is possible that tidal heating of ancient infall UDGs might become `too efficient' in this region, instead leading to compact galaxies from tidal stripping, complete disruption or the formation of ultra-compact dwarf galaxies from nucleated UDGs \citep{Janssens_et_al_19, Wang_et_al_23}.
In this interpretation, GC-poor UDGs in the ancient infall region of phase-space might instead have more recent infall times (\citealt{Rhee_et_al_17} find that nearly 50 per cent of galaxies in the `ancient infall' region of phase-space actually have later infall redshifts).

\begin{figure}
    \includegraphics[width=\columnwidth]{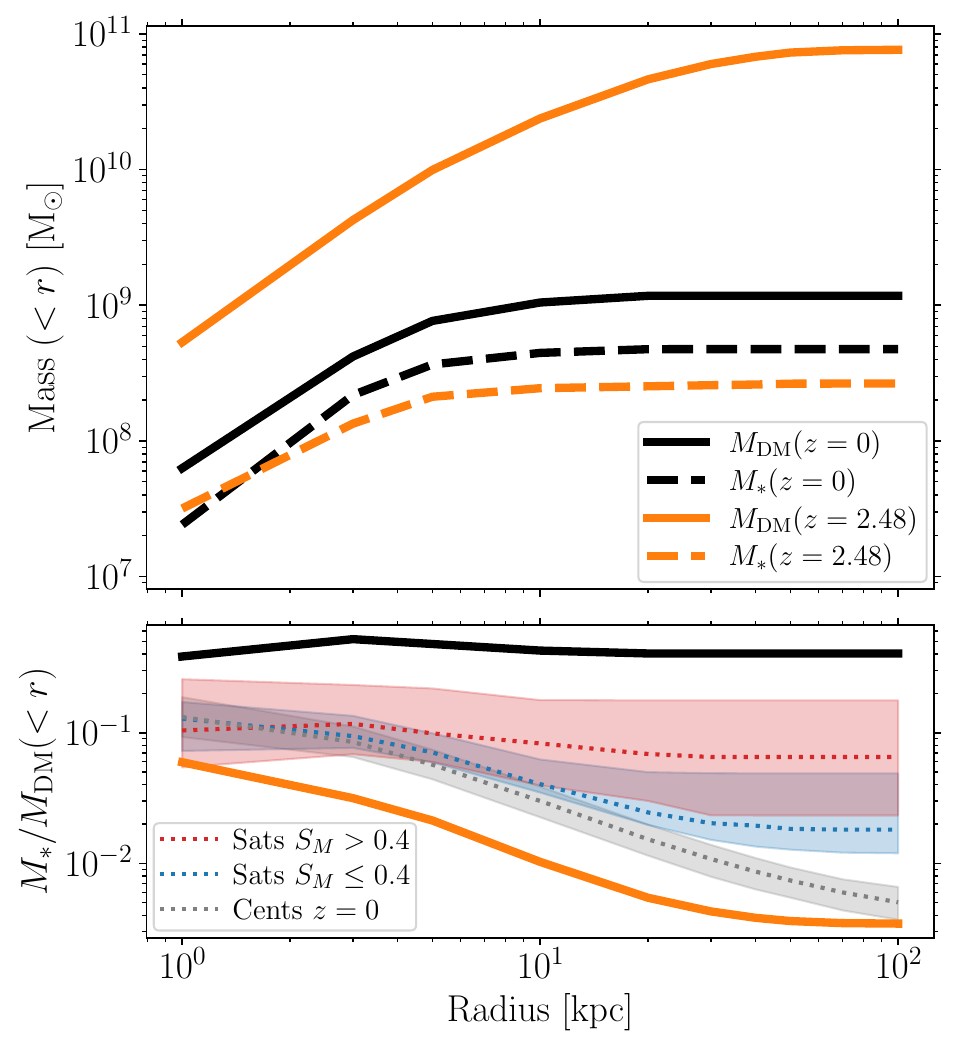}
    \caption{Comparison of the initial (redshift prior to becoming a satellite, thick orange lines) and $z=0$ (thick black lines) enclosed mass profiles for a galaxy which has undergone strong tidal stripping of the dark matter halo (i.e.\ highest $M_\ast / M_\mathrm{dyn}$ in Figure~\ref{fig:Mratio}). In the upper panel, solid lines show the dark matter mass profiles, while dashed lines show the stellar mass profiles. The lower panel shows the stellar-to-dark matter mass profiles at each redshift. The mass profile is shown at larger radii than the gravitational softening length ($0.35 \kpc$). Unlike the initial profile, at $z=0$ the stellar-to-dark matter mass ratio is roughly constant ($\approx 0.5$). For comparison, the dotted grey, blue and red lines show the median mass ratio at each radius for $z=0$ central galaxies and GC-poor ($S_M \leq 0.4$) and GC-rich ($S_M > 0.4$) satellite galaxies in groups/clusters (i.e. those in Figure~\ref{fig:Mdyn}), respectively, with shaded regions showing the 16\ts{th} to 84\ts{th} percentiles for each population. As found in Figures~\ref{fig:Mdyn} and \ref{fig:Mratio}, from GC-poor to GC-rich satellite galaxies there is an increasing trend of dark matter mass loss compared to central galaxies.}
    \label{fig:profile}
\end{figure}

The enclosed dynamical masses of galaxies are, of course, also dependent on their dark matter profiles.
Many works suggest that UDGs reside within cored, rather than cuspy \citep[i.e.\ an NFW profile,][]{Navarro_Frenk_and_White_96, Navarro_Frenk_and_White_97}, dark matter haloes \citep[e.g.][]{Di_Cintio_et_al_17, Carleton_et_al_19, Martin_et_al_19, van_Dokkum_et_al_19}, where the cores are generally thought to form through stellar feedback processes \citep[e.g.][]{Navarro_Eke_and_Frenk_96, Read_and_Gilmore_05, Governato_et_al_10, Maccio_et_al_12, Pontzen_and_Governato_12}.
Such a feedback-driven expansion is also potentially more efficient in GC-rich galaxies \citep[e.g.][]{Trujillo-Gomez_et_al_21,Trujillo-Gomez_et_al_22}.
Galaxies in the EAGLE model (and thus also in this work) do not form dark matter cores unless the star formation density threshold is significantly increased \citep{Benitez-Llambay_et_al_19}.
However, EAGLE galaxies may appear cored in mock observations \citep{Oman_et_al_19} and models with cusps may better explain the diversity of galaxy rotation curves than those with cores \citep{Roper_et_al_23}.

The measured kinematic profiles of UDGs are contradictory: some UDGs are consistent with core profiles \citep{van_Dokkum_et_al_19, Gannon_et_al_22, Gannon_et_al_23}, and others are consistent with NFW haloes \citep[though see also \citealt{Brook_et_al_21} who suggest a shallower inner profile slope in the case of AGC 242019]{Leisman_et_al_17, Sengupta_et_al_19, Shi_et_al_21}.
In general, the dynamics of cluster and field galaxies are determined using different methods (using stellar and gas kinematics, respectively) and thus have different sources of uncertainties and systematics.
In the case of gas rotation profiles, many types of perturbations can affect their dynamics, leading to rotation profiles that do not match the true circular velocity profiles \citep{Downing_and_Oman_23} and cuspy halo profiles appearing to be cored \citep{Hayashi_and_Navarro_06, Pineda_et_al_17, Oman_et_al_19, Roper_et_al_23}.
From their stellar dynamics, the cluster UDGs tend to favour cored profiles \citep{van_Dokkum_et_al_19, Gannon_et_al_22, Gannon_et_al_23} when assuming standard mass--concentration relations for cuspy haloes.
However, as we find in Figure~\ref{fig:Mdyn}, inferring the initial dark matter halo profile for satellite galaxies in clusters from their dynamical masses may be difficult, as most galaxies (except for very recent accretion) have undergone significant tidal stripping of their dark matter halo.
In this case, cored dark matter profiles could be inferred from lower-than-expected dynamical masses, despite having cuspy profiles \citep[cusps are retained during tidal stripping,][]{Kazantzidis_et_al_04}.

In Figure~\ref{fig:profile} we compare the stellar and dark matter mass profiles for one of the simulated galaxies in Figure~\ref{fig:Mratio} that have undergone the most tidal stripping (i.e.\ have the highest $M_* / M_\mathrm{dyn}$; the result for the other galaxy is similar).
At $z=2.48$, prior to becoming a satellite, the dark matter halo of the galaxy initially follows an NFW profile (solid orange line in top panel), though the smaller virial radius of the high redshift halo compared to typical $z=0$ haloes (see also Section~\ref{sec:SM-Mdm}) results in a lower stellar-to-dark matter mass ratio at small radii (bottom panel, orange line compared to grey dotted line).
However, we find that the $z=0$ dark matter profile of the galaxy appears approximately as scaled version of the stellar mass profile (the enclosed stellar mass-to-dark matter mass ratio is nearly constant with radius) rather than an NFW profile \citep[which might be expected for tidally-limited galaxies,][]{Errani_et_al_22}.
Thus, galaxies in such highly-stripped haloes will have very different mass profiles compared to galaxies residing in typical isolated dark matter haloes at $z=0$ (grey line in the figure), which should be taken into account when fitting dynamical mass profiles to observed galaxies.
In the median, the differences in enclosed stellar-to-dark matter ratios compared to $z=0$ central galaxies increases from GC-poor satellites (blue dotted line) to GC-rich satellites (red dotted line) at radii $\gtrsim 5 \kpc$ due to increased tidal stripping of the dark matter haloes.

\section{Summary}
\label{sec:summary}

In this work, we used simulations of galaxies and their GC systems from the E-MOSAICS project to investigate possible origins of the observed correlation between stellar velocity dispersion (or dynamical mass) and GC system richness in UDGs \citep{Gannon_et_al_22}.
We found that the simulations in fact predict the opposite trend, with GC-rich galaxies having lower enclosed masses than GC-poor galaxies (Section~\ref{sec:infall_times}).
This occurs due to the earlier infall times ($z_\mathrm{infall} \gtrsim 1$) of GC-rich galaxies in galaxy groups/clusters, resulting in increased tidal stripping of their dark matter haloes compared to GC-poor galaxies.
However, the enclosed enclosed masses for the simulated GC-rich galaxies agree well with the dynamical masses of observed GC-rich UDGs.
Therefore, we find that GC-rich UDGs are consistent with being a population of early-infalling galaxies \citep[see also][]{Carleton_et_al_21} which have subsequently undergone tidal stripping of much of their dark matter haloes (a process that could also result in tidal heating to explain the large sizes of UDGs, e.g. \citealt{Yozin_and_Bekki_15, Safarzadeh_and_Scannapieco_17, Ogiya_18, Carleton_et_al_19, Sales_et_al_20}; though this may be at odds with the phase-space locations of GC-rich UDGs in clusters, \citealt{Forbes_et_al_23}, see Section~\ref{sec:UDG_formation} for further discussion).

In contrast, the simulated GC-poor galaxies have enclosed masses that are generally inconsistent with those found for GC-poor cluster UDGs from their stellar velocity dispersions \citep{Gannon_et_al_22}.
However, the enclosed masses of isolated and late infalling ($z_\mathrm{infall} \lesssim 0.3$) simulated galaxies do agree well with the dynamical masses of HI-rich field galaxies and edge-on UDGs \citep{Lelli_et_al_16, He_et_al_19}.
If tidal stripping of the dark matter haloes were to explain the lower inferred dynamical masses of GC-poor cluster UDGs, it would require GC-poor galaxies to have earlier infall times or be preferentially more affected than GC-rich galaxies, which we consider to be unlikely.
Instead, we considered whether the kinematics of GC-poor galaxies are systematically different from those of GC-rich galaxies.
If GC-poor UDGs are rotating systems observed nearly face on, such that their velocity dispersion is only representative of the vertical disc support, their velocity dispersions would underestimate the total mass in the system.
This may be particularly important for higher mass UDGs (i.e.\ typical spectroscopic targets), as edge-on systems may not be classified as UDGs due to their higher surface brightnesses \citep{He_et_al_19, Van_Nest_et_al_22}.

In this work, we used the simulations to show that galactic conditions for star formation could result in an anti-correlation between the rotational support of galaxies and their GC system richness (Sections~\ref{sec:dynamics} and \ref{sec:origin}).
Galaxies with higher gas fractions are turbulent systems with low rotational support \citep[e.g.][]{Dekel_et_al_09, Genzel_et_al_11}, but typically have high gas pressures (Figure~\ref{fig:fgas}), leading to efficient GC formation \citep[Figure~\ref{fig:SM-P},][]{Kruijssen_12, P18}.
Conversely, lower gas fractions lead to more rotational support and lower gas pressures, resulting in less efficient GC formation.
Though we focus on a small galaxy mass range in this work ($2 \times 10^8 < M_\ast / \Msun < 6 \times 10^8 \Msun$), such processes are not unique to any mass range, and may therefore explain why early-type galaxies at all masses typically have richer GC systems than late-type galaxies \citep{Georgiev_et_al_10}.

Though there are currently no GC-poor UDGs in clusters with resolved stellar kinematic profiles to test for rotation, oblate rotating galaxies would be expected to have a different distribution of shapes (projected axis ratios, $b/a$) to dispersion-supported spheroidal galaxies.
For UDGs in the Coma cluster, galaxies with lower GC specific frequencies are indeed found to have lower axis ratios than those with high specific frequencies \citep{Lim_et_al_18}, which may indicate that they are (nearly) oblate systems.
We expanded this comparison to include Virgo cluster dwarf galaxies, UDGs in the Perseus cluster, and UDGs from the MATLAS survey (Section~\ref{sec:comparison}).
In all observed galaxy samples, GC-rich galaxies are found to have relatively round shapes ($b/a \gtrsim 0.6$) that are similar to those of elliptical galaxies, while GC-poor galaxies typically have a wider range of shapes ($b/a \approx 0.3$-$1$) that are similar to spiral galaxies.
No very flattened galaxies ($b/a < 0.5$) were found to be GC-rich systems.

Therefore, current observations support a scenario where GC-poor galaxies are (on average) discy, rotating systems.
This may explain the low observed stellar velocity dispersions of GC-poor UDGs in the Perseus cluster \citep{Gannon_et_al_22}, implying underestimated dynamical masses if they are near face-on discs, and reconciling their masses with those of simulated galaxies (Section~\ref{sec:infall_times}).
This would also make GC-poor UDGs natural analogues of HI-rich field UDGs, which have since lost their gas after entering the group/cluster environment.
Future observations should test directly for our predicted stellar rotation within GC-poor UDGs, as well as further dynamical mass measurements of highly flattened UDGs, where inclination corrections are small.

\section*{Acknowledgements}

This work was supported by the Australian government through the Australian Research Council's Discovery Projects funding scheme (DP220101863) and the Australian Research Council Centre of Excellence for All Sky Astrophysics in 3 Dimensions (ASTRO 3D), through project number CE170100013.
JMDK gratefully acknowledges funding from the European Research Council (ERC) under the European Union's Horizon 2020 research and innovation programme via the ERC Starting Grant MUSTANG (grant agreement number 714907). COOL Research DAO is a Decentralized Autonomous Organization supporting research in astrophysics aimed at uncovering our cosmic origins.
AJR was supported by National Science Foundation grant AST-2308390.  Support for Program number HST-GO-15235 was provided through a grant from the STScI under NASA contract NAS5-26555.
This work used the DiRAC Data Centric system at Durham University, operated by the Institute for Computational Cosmology on behalf of the STFC DiRAC HPC Facility (\url{www.dirac.ac.uk}). This equipment was funded by BIS National E-infrastructure capital grant ST/K00042X/1, STFC capital grants ST/H008519/1 and ST/K00087X/1, STFC DiRAC Operations grant ST/K003267/1 and Durham University. DiRAC is part of the National E-Infrastructure.
The work also made use of high performance computing facilities at Liverpool John Moores University, partly funded by the Royal Society and LJMU's Faculty of Engineering and Technology.

\section*{Data Availability}

The data underlying this article will be shared on reasonable request to the corresponding author.


\bibliographystyle{mnras}
\bibliography{emosaics}



%
%


\bsp	
\label{lastpage}
\end{document}